\documentclass[twocolumn]{aastex631}
\newcommand{\ba}{\begin{eqnarray}}
\newcommand{\ea}{\end{eqnarray}}

\usepackage{lineno}
\shorttitle{Disk imprints from multiplanet systems}
\shortauthors{Garrido-Deutelmoser et al. }

\graphicspath{{./}{figures/}}

\begin{document}

\title{Substructures in protoplanetary disks imprinted by compact planetary systems}

\author[0000-0002-7056-3226]{Juan Garrido-Deutelmoser}
\affiliation{Instituto de Astrofísica, Pontificia Universidad Católica de Chile, Av. Vicuña Mackenna 4860, 782-0436 Macul, Santiago, Chile.}
\affiliation{N\'ucleo Milenio de Formaci\'on Planetaria (NPF), Chile.}

\author[0000-0003-0412-9314]{Cristobal Petrovich}
\affiliation{Instituto de Astrofísica, Pontificia Universidad Católica de Chile, Av. Vicuña Mackenna 4860, 782-0436 Macul, Santiago, Chile.}
\affiliation{Millennium Institute for Astrophysics, Chile.}

\author[0000-0001-7671-9992]{Leonardo Krapp}
\affiliation{Department of Astronomy and Steward Observatory, University of Arizona, Tucson, Arizona 85721, USA}

\author[0000-0001-5253-1338]{Kaitlin M. Kratter}
\affiliation{Department of Astronomy and Steward Observatory, University of Arizona, Tucson, Arizona 85721, USA}

\author[0000-0001-9290-7846]{Ruobing Dong}
\affiliation{Department of Physics and Astronomy, University of Victoria, Victoria, British Columbia, Canada.}

\begin{abstract}
The substructures observed in protoplanetary disks may be the signposts of embedded planets carving gaps or creating vortices. The inferred masses of these planets often fall in the Jovian regime despite their low abundance compared to lower-mass planets, partly because previous works often assume that a single substructure (a gap or vortex) is caused by a single planet. In this work, we study the possible imprints of compact systems composed of Neptune-like planets ($\sim10-30\;M_\oplus$) and show that long-standing vortices are a prevalent outcome when their inter-planetary separation ($\Delta a$) falls below $\sim8$ times $H_{{\rm p}}$---the average disk's scale height at the planets locations. In simulations where a single planet is unable to produce long-lived vortices, two-planet systems can preserve them for at least $5,000$ orbits in two regimes: i) fully-shared density gaps with elongated vortices around the stable Lagrange points $L_4$ and $L_5$ for the most compact planet pairs ($\Delta a \lesssim 4.6\; H_{{\rm p}}$); ii) partially-shared gaps for more widely spaced planets ($\Delta a \sim 4.6 - 8\;H_{{\rm p}}$) forming vortices in a density ring between the planets through the Rossby wave instability. The latter case can produce vortices with a wide range of aspect ratios down to $\sim3$ and can occur for planets captured into the 3:2 (2:1) mean-motion resonances for disk's aspects ratios of $h\gtrsim 0.033$ ($h\gtrsim 0.057$). We suggest that their long lifetimes are sustained by the interaction of spiral density waves launched by the neighboring planets. Overall, our results show that distinguishing imprint of compact systems with Neptune-mass planets are long-lived vortices {\it inside} the density gaps, which in turn are shallower than single-planet gaps for a fixed gap width.
\end{abstract}


\section{Introduction}
\label{sec:Intro}

Substructures in protoplanetary discs (PPDs) such as rings, gaps, spirals, vortices, and other non-axisymmetric substructures are often shown by high angular resolution observation in sub-millimeter/millimeter and optical/near-infrared wavelengths \citep{ALMA2015,Andrews2020,Francis2020}. Although several formation mechanisms have been proposed, their origin remains still unclear. These mechanisms may involve either internal processes within the disks such as hydrodynamic instabilities \citep{Takahashi2016,Takahiro2021,Pierens2021}, magneto-hydrodynamic effects \citep{Flock2015,Krapp2018}, the presence of snowlines \citep{Zhang2015,Okuzumi2016}, among others, or may be driven by planets through disk-planet interactions \citep{Goldreich1980,LinDNC1986}. 

Interactions with planets seem to reproduce at least some observed substructures  \citep[e.g.,][]{Pinilla2012,Dipierro2015,Dong2015,Zhang2018,Hammer2021,Rodenkirch2021,Fedele2021}, including the remarkable case of PDS70 where such planets have been observed \citep{Keppler2018,Muller2018,Wagner2018,Haffert2019}. As in PDS70, super Jovian-mass planets are often invoked to reproduce the dust continuum observations by ALMA \citep{Long2018,Bae2018,Lodato2019}, though such  planets have small occurrence rates as shown by direct imaging campaigns \citep{Bowler2018}; a potential tension that is alleviated if the planets eventually migrate inwards \citep{MarelMulders}. 

Lower-mass planets ($M_{{\rm p}} \lesssim 30\; M_{\rm \oplus}$) are observed to be much more common than the Jovians, at least inside $\sim 10$ au as shown by microlensing detections \citep[see][Fig. 21]{Zhang2018} and radial velocity campaigns \citep{Cumming2008,Mayor2011}. At closer-in distances to their host stars ($\lesssim1$ au), the NASA Kepler Space Telescope revealed this overabundance of sub-Neptune planets pursues, possibly outnumbering the stars themselves \citep{Petigura2013}. A common feature of these planets that they belong to multi-planet systems (average of $\sim 3$ planets in inside $\sim 1$ au; \citealt{Zhu2018}), sometimes in very compact configurations \citep{Lissauer2011a,Lissauer2014,Rowe2014}, ranging from two-planet systems such as Kepler-36 \citep{Carter2012} to high-multiplicity ones like Kepler-11 \citep{Lissauer2011b}. However, Kepler systems may not be directly related to the structures observed by ALMA, as they differ in orbital radii by a factor of a few tens.

Despite the popularity of this planetary population in our Galaxy, including the pair Uranus-Neptune in our Solar System, the question of what possible imprints they may leave in the disk remains poorly studied. Instead, most works have focused on the effects that disk has on the planetary dynamics and their migration \citep{Broz2018,Cimerman2018,Kanagawa2020,Cui2021}. In this paper, we fill this gap by studying the different features and structures in PPDs imprinted by sub-Jovian mass planets in compact configurations. 

\paragraph{Disk-planet interactions: role of multiplicity}

The effect on multi-planet systems on PPDs has been explored as a mechanism to open wide cavities using several Jovian-sized planets, offering an explanation to the observed transition disks \citep{Zhu2011}. An example of a wide gap is that is of PDS70, where the carving of gas and dust is well-explained by two super-Jupiters sharing a common gap \citep{Bae2019}. Given our exhaustive knowledge of gaps opened by single planets  \citep[e.g.,][]{Kanagawa2015,Duffell2020}, one may be tempted to model the cavities in multi-planet systems as the {\it superposition} of individual gaps centered around each planet. However, as shown by \citet{DuffellDong2015} multiple planets in a common cavity produce shallower gaps than single planets as a result of angular momentum transfer, so the superposition would not be applicable.\\

These works have mainly focused on a regime where $\alpha \sim 10^{-3} - 10^{-2}$ and the planets have Jovian masses, often super-thermal. At lower viscosities, planets with masses even below the thermal mass, defined as
\begin{eqnarray}
    M_{\rm th} = \frac{c_{\rm s}^3}{G \Omega} = h_{{\rm p}}^3 M_{\rm \star},
\label{eq:Mth}
\end{eqnarray}
can open gaps or even multiple gaps \citep{Goodman2001}. Moreover, such planet may still produce non-axisymmetric structures such as vortices \citep[e.g.,][]{Hammer2021} or over-densities around the Lagrange points $L_{\rm 4}$ and $L_5$ \citep[e.g.,][]{Rodenkirch2021} that may add complexity to the morphologies of PPDs when multiple planets are present. Our work aims at understanding this potential complexity.

Our paper is organized as follows. We describe the setup and details of our numerical simulations in Section \ref{sec:2_Set-up}. We present our main results in \S \ref{sec:3_Results}, which we classify in different regimes in \S\ref{section:4_Classification_Regimes}. We provide with more details and a physical interpretation of two relevant regimes in \S\ref{sec:5_RegimesII} and \S \ref{sec:6_RegimesIII}. In \S\ref{section:7_Discussion} we discuss our main results and observational implications, and end with our conclusions in \S \ref{section:8_Conclutions}.

\section{Set-Up}
\label{sec:2_Set-up}

We carried out 2D hydrodynamic simulations using \textsc{Fargo3D} code \citep{Benitez2016,Benitez2019,Masset2000} that solves Navier-Stokes equations, considering vertically integrated quantities. These equations are the conservation of mass
\begin{equation}
    \frac{\partial \Sigma}{\partial t} + \nabla \cdot (\Sigma \vec{v})= 0,
\label{eq:cons_mass}
\end{equation}
and the conservation of momentum
\begin{equation}
    \Sigma \left( \frac{\partial \vec{v}}{\partial t}+\vec{v} \cdot \nabla \vec{v} \right) = -\nabla P -\Sigma\nabla \vec{\phi} + \nabla \cdot \vec{T},
\label{eq:cons_momentum}
\end{equation} 
where $\Sigma$ is the surface density, $\vec{v}$ the flow velocity vector, $\vec{\phi}$ the gravitational potential of the star and the planets, $\vec{T}$ the viscous stress tensor, and $P$ the pressure. We assume a locally isothermal equation of state $P = \Sigma c_{\rm s}^{2}$ where $c_{\rm s} = H \Omega_{\rm K}$ is the sound speed, $\Omega_{\rm K}$ the local Keplerian orbital frequency and $H$ the scale height of the disk.

The planets are allowed to move freely, interacting gravitationally with other planets and the gaseous disk.

\subsection{Planetary systems}

We define the spacing of the planets as
\begin{equation}
\Delta a\equiv a_{{\rm p},i+1}-a_{{\rm p},i},
\end{equation}
which we express in units the mutual Hill radius given by
\begin{equation}
    R_{\rm H}=\frac{a_{{\rm p},i}+a_{{\rm p},i+1}}{2}\left(\frac{m_{{\rm p},i}+m_{{\rm p},i+1}}{3M_\star}\right)^{1/3},
\label{eq:Rhill}
\end{equation}
where $m_{{\rm p},i}$ is the mass of the planet $i$ orbiting a Sun-like star ($M_\star = 1\;M_\odot$) at a semi-major axis $a_{{\rm p},i}$ ($a_{{\rm p},i}<a_{{\rm p},i+1}$).

We consider a range of $\Delta a\sim 4 - 11\;R_{\rm H}$. Theoretically, such compact systems may emerge from oligarchic growth, where proto-planets end at orbital separations in the range $\Delta a= 5 - 10\;R_{\rm H}$  \citep{Kokubo98}. While observationally, they may be precursors of Kepler-like with typical $\Delta a\sim 10-20\;R_{\rm H}$ previous to the last mass-doubling giant impacts \citep[e.g.,][]{PuWu2015,Yee2021}.

\subsection{Fiducial model}

We choose a power-law profile for the initial disk surface density $\Sigma_0 = 1.7\;(r/r_0)^{-1.5} \;\rm gr/cm^2$, corresponding to $0.1\%$ of the MMSN \citep{Hayashi1981}. Similar for the sound speed $c_{\rm s} \propto r^{-1/2}(r/r_0)^{f}$ using the flaring index $f=1/7$ and $r_0 = a_i = 1$ au, corresponding to a $c_{\rm s} \sim$ $1.53\;\rm km/s$ at $1$ au. 
\begin{itemize}
    \item  The disk extent from $r_{\rm in} = 0.3$ au to $r_{\rm out} = 4.0$ au, implying an initial disk mass of $\sim 1 \; M_{\rm \oplus}$, making self-gravity and planet migration\footnote{According to \citet[][Eq. 5,]{Kanagawa2018} the migration timescale correspond to $\tau_{\rm a} \sim 10^{7}$ orbits for the fiducial model.} negligible. The computational domain is composed of $n_{\rm r} = 768$ logarithmically-spaced radial cells and $n_{\rm \theta} = 1280$ equally-spaced cells in the azimuthal $[0,2\pi]$ domain. 
    
    \item  The standard viscosity parameter $\alpha$ \citep{Shakura1973} is set as a constant throughout the disk with $\alpha = 10^{-4}$. 
    \item An aspect ratio of  $h(r) = h_0(r/r_0)^f$ with $h_0=0.0516$
     \item An inner planet at $r_{{\rm p},i} = 1$ au, while the outer planet is place at $\Delta a =4.3 - 10.9\;R_{\rm H}$ from the inner planet (Table \ref{tab:models}). Both planets start with low eccentricities of $e=0.005$ and their final masses\footnote{We tested simulations gradually growing the planets and find similar results.}.
\end{itemize}
Finally, damping regions for the radial boundaries are applied at a distance of $ 1.53$ ($1/1.53$) times that of the inner (outer) boundary radius \citep{deValBoro2006}. The various planetary separations $\Delta a$ and masses $m_{{\rm p}}$ as well as other disk parameters ($h_0$ and $n_{\rm r} \times n_{\rm \theta}$), different from the fiducial values above, are shown in Table \ref{tab:models}.

\section{Results}
\label{sec:3_Results}

\defcitealias{Duffell2020}{D20}
\defcitealias{Kanagawa2017}{K17}
\defcitealias{Kanagawa2015}{K15}

\begin{figure*}[t]
    \centering
    \includegraphics[width=0.99\linewidth]{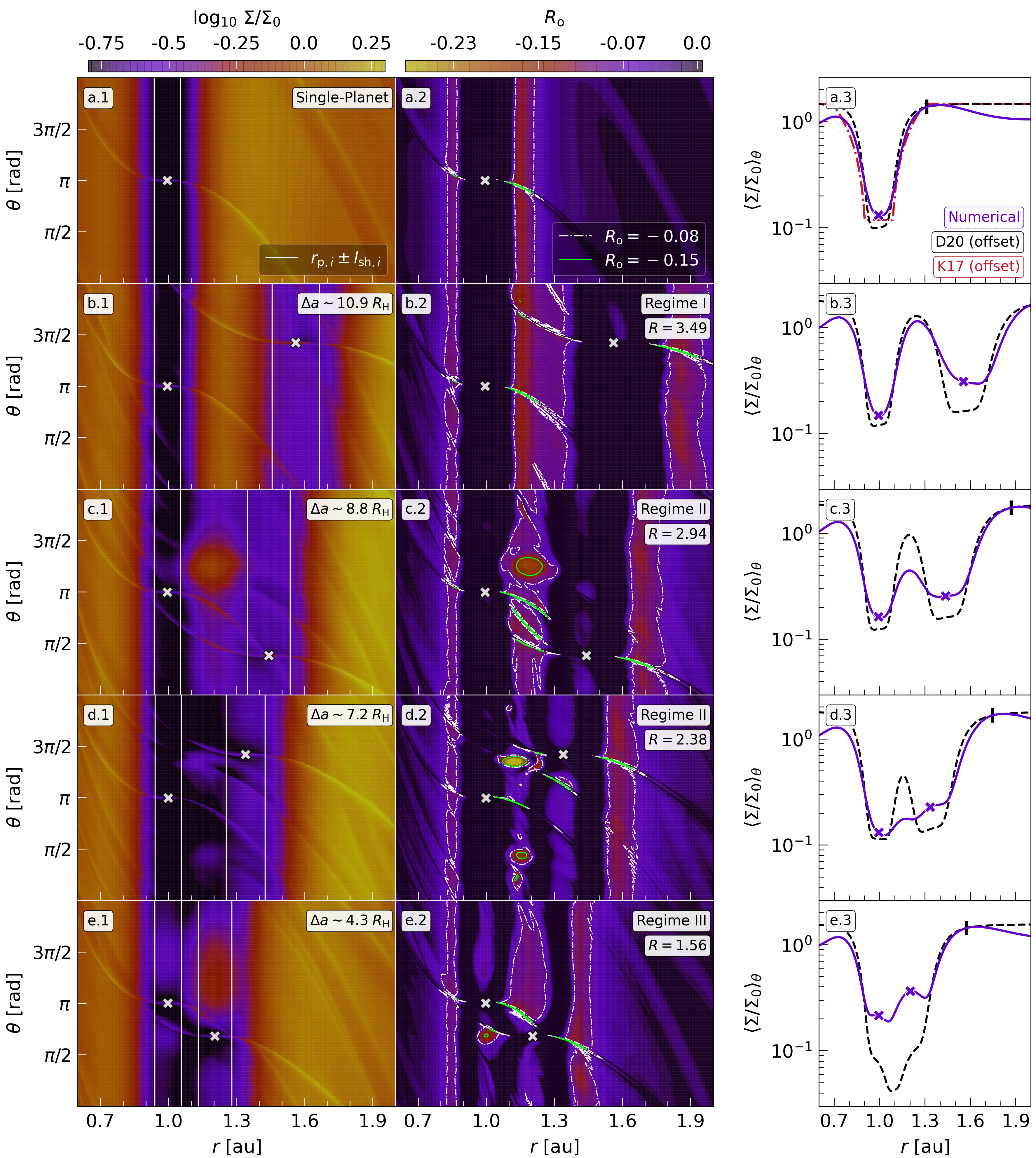}
    \caption{Columns show snapshots after 5000 orbits of the inner planet. Left: Maps $\Sigma/\Sigma_0$ in log-scale. The white vertical lines indicate the shock length $\pm l_{\rm sh}$ (Eq. \ref{eq:lsh}) from each planet, where the crosses denotes their position. Middle: Contours of the Rossby number $R_{\rm o}$, constrained to negative values displaying anticyclonic rotation. White and green, contours denote $R_{\rm o} = {-0.08,-0.15}$. Right: Azimuthally-averaged $\Sigma/\Sigma_0$ profiles. The red and black dashed lines indicate the analytic profiles provided by \citetalias{Kanagawa2017} and \citetalias{Duffell2020} for a single planet, where we apply a multiplicative factor to these profiles to match the simulations at the outer edge, location indicated as a vertical black tick. Rows correspond to different models with a single-planet (a) ${\rm M1\_\Delta0.0}$, and two-planet models (b) ${\rm M1\_\Delta10.9}$, (c) ${\rm M1\_\Delta8.8}$, (d) ${\rm M1\_\Delta7.2}$, and (e) ${\rm M1\_\Delta4.3}$, see Table \ref{tab:models} for details of the simulations.} 
    \label{fig:Main_Panels}
\end{figure*}

In Figure \ref{fig:Main_Panels} we show the simulation results for a single-planet (panels a.1 - a.3) and two-planet models with separations of $\Delta a\sim 10.9$, $8.8$, $7.2$, and $4.3\;R_{\rm H}$ (panels b.1 - e.3, see Table \ref{tab:models}) after $5,000$ inner planetary orbits. In what follows, we navigate the results going from the left to the right panels.

\subsection{Disk morphology (left panels)}

The panel (a.1) shows that a single planet opens a gap with a fully axi-symmetric density distribution and no co-rotational accumulations. The small departures are due to the spiral density waves launched by the planet. Similar to this single-planet case, panel (b.1) with two well-separated planets also displays only axi-symmetric structures, including now a ring that separates both planets. This is expected for significant inter-planetary separations since each planet carves its gap nearly unaffected by its neighbor.

As the planetary separations drop below $\Delta a \sim10 \;R_{\rm H}$ (panels c.1, d.1, and e.1) we observe the appearance of non-axisymmetric structures. Panel (c.1) shows planets separated by a narrow ring, trapping gas and forming an elongated vortex-like sub-structure. Panel (d.1) shows planets partially sharing their gaps separated by tenuous and compact substructures. Finally, panel (e.1) shows gas trapped in the co-orbital region for each planet, displaying over-densities at the Lagrange points $L_4$ and $L_5$, which are located in the leading and trailing position respect to the planet at $\pm\pi/3$.

\subsection{Vortices (middle panels)}

A useful quantity to measure the strength of the vortices is provided by the Rossby number, defined as
\begin{equation}
    R_{\rm o} = \frac{\nabla \times (v-v_{\rm K})_{\rm z}}{2\Omega_{\rm K}},
\label{eq:R0}
\end{equation}
where $v_{\rm K} = \Omega_{\rm K} r$ is the Keplerian orbital velocity.

A practical classification of vortices is provided by \citet[][Sec. 5.4]{Surville2015}. The authors identify two families: \textit{Incompressibles} for which $R_{\rm o} > -0.15$ with an elongated shape and \textit{compressibles} for which $R_{\rm o} < -0.15$ with a compact shape. Their streamlines can break the flow around the vortex by exciting spiral waves from the vortex core. 

Despite the fact that the gap edges do show regions with $R_{\rm o}\sim -0.1$, no vortex survives after $5,000$ orbits in our simulations in these regions. In turn, Panel (c.2) shows an elongated vortex in between the planets that reaches $R_{\rm o} \sim -0.16$. Similarly, Panel (d.2) shows compact vortices with the strongest one ($R_{\rm o} \lesssim -0.27$) launching spirals from their core. Finally, panel (e.2) shows the expected anticyclonic motion ($R_{\rm o}<0$) at $L_4$ and $L_5$, characterized by tadpole orbits around these equilibrium points.

\subsection{Azimuthally-averaged density profiles (right panels)}

We compare our numerical azimuthally-averaged density profiles with analytic expressions obtained in single-planet calculations by  \citet[Eq. 6,][hereafter K17]{Kanagawa2017} and \citet[Eq. 19,][hereafter D20]{Duffell2020}. A multiplicative factor is applied to match the profiles at the outer edge of the gap, due to the pile-up produced in our simulations, expected in low-viscosity and low-mass disks  \citep{Dempsey2020}.
For a single planet (panel a.3), we show that \citetalias{Kanagawa2017} and \citetalias{Duffell2020} reproduce the numerical profile reasonably well, but with \citetalias{Duffell2020} predicting slightly deeper gaps than observed. 

For the two-planet cases (panels b.3 - e.3) we only show the profile by \citetalias{Duffell2020}. Overall, we find that it can reproduce gap widths reasonably well, but the simple addition of two independent profiles does not capture the genuine behavior of gap opening by two planets, both in depths at planet positions, as well as the amount of material between the planets (bumps in panels c.3 and d.3). In the most compact two-planets systems in panel (e.3), the gap depth is shallower by a factor of $2$ compared to the single-planet case. This behavior had been observed in the simulations by \citet{DuffellDong2015} for cavity-sharing Jovian-mass planets and attributed to the cavity refilling from the density waves launched by the planetary  neighbors. In our case, the gaps are also shallower due to the gas accumulation at $L_4$ and $L_5$. We discuss this further in \S\ref{sec:6_RegimesIII}.

\section{Classification of regimes} \label{section:4_Classification_Regimes}

In this section we attempt to provide with a set of criteria to classify the different set of morphologies observed in Figure \ref{fig:Main_Panels} based on the disk properties and planetary separations. Before doing this, we revise some general aspects of spiral density wave damping and vorticity excitation that will allow us to perform a more quantitative classification.

\subsection{Spiral density waves: shocks and vorticity}\label{subsection:context}

A planet embedded in a PPD excites density waves at Lindblad resonances, which propagate across the disk carrying energy and angular momentum. Its deposition in the disk, which depends on the dissipation mechanism, may give rise to gas evolution and the formation of substructures. In particular, \citet{Goodman2001} showed these waves may damp non-linearly by wave steeping and subsequent shocks, even for sub-thermal mass planets in a low viscosity environment. The typical distance from the planet at which the wave shocks is given by
\begin{equation}
    l_{\rm sh} \approx 0.93H(r_{{\rm p}})\left(\frac{\gamma+1}{12/5}\frac{M_{{\rm p}}}{M_{\rm th}}\right)^{-2/5},
\label{eq:lsh}
\end{equation}
while the angular momentum flux (AMF) transported by the spiral, decays in the post-shock region \citep{Goodman2001,Rafivok2002a,Dong2011a} as
\begin{equation}
    \textrm{AMF}(r) \propto |r-r_{{\rm p}}|^{-5/4} \quad  (|r-r_{{\rm p}}| \gg l_{\rm sh}).
    \label{eq:AMF}
\end{equation}

As for the vortices, these can also be generated by planets through the Rossby Wave Instability \citep[RWI:][]{Lovelace1999,Li2000,Li2001}.  The critical function for the RWI in a locally isothermal disk is given by \citet{Lin2012} as $L_{\rm  iso}= c_{\rm s}^2/\zeta$, which depends inversely on the vortensity 
\begin{equation}
    \zeta = \frac{(\nabla \times v)_z}{\Sigma}.
\end{equation}

The RWI can be triggered in regions where $L_{\rm iso}$ becomes maximum, such as the gap edges. In turn, the vortensity perturbation $\Delta\zeta = \zeta - \zeta_0$ produced by spiral shocks is a steep function of the planet mass, where the peak scales as $\Delta\zeta_{{\rm peak}} \propto (M_{{\rm p}}/M_{\rm th})^{2.95}$ \citep{Dong2011a}. Recently, \citet{Cimerman2021} have shown that $\Delta\zeta_{{\rm peak}}$ is produced between $1 \lesssim |r-r_{{\rm p}}|/l_{\rm sh} \lesssim 2$, with an exact location that depends on $M_{{\rm p}}/M_{\rm th}$. The vortensity peak is followed by a negative perturbation or a maximum in $L_{\rm iso}$, which locates a region where the gas could become unstable to the RWI under favorable conditions, e.g., sufficiently sharp density profile \citep{Ono2016} and low viscosity ($\alpha < 10^{-3}$) \citep{deValBorro2007}.

\begin{figure}[htp]
    \centering
    \includegraphics[width=0.99\linewidth]{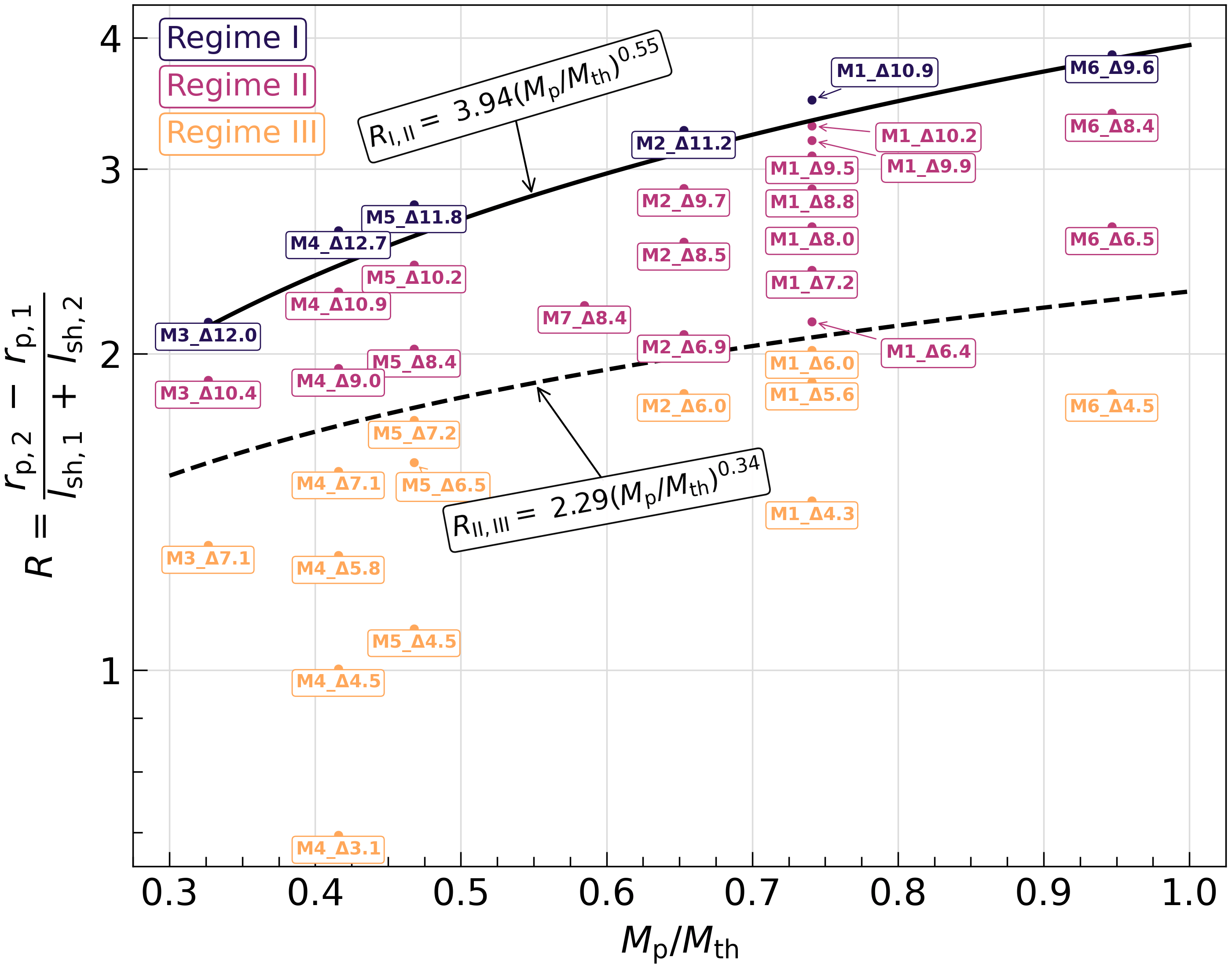}
    \caption{Classification of the regimes, I ({\it non-overlapping gaps and no substructure}), II ({\it substructure between planets}) and  III ({\it common gap and sub-structure at Lagrange points}), colors were assigned by visual inspection following the description in \S\ref{subsec:Regimes}, as a function of $M_{{\rm p}}/M_{\rm th}$ and $R=(r_{{\rm p},2}-r_{{\rm p},1})/(l_{\rm sh,1}+l_{\rm sh,2})$ (Eq. \ref{eq:R}). Solid and dashed black lines delimit the regimes I/II and II/III by $R_{\rm \rm{I,II}} = 3.94(M_{{\rm p}}/M_{\rm th})^{0.55}$ and $R_{\rm \rm{II,III}}  = 2.29(M_{{\rm p}}/M_{\rm th})^{0.34}$ respectively.}
    \label{fig:Regimes_Plot}
\end{figure}

\subsection{Proposed regimes}
\label{subsec:Regimes}

Hinted by the vertical lines of Figure \ref{fig:Main_Panels}, showing that $\pm  l_{\rm sh} $  encloses well the radial extent of the density features (ring and vortices), we define the dimensionless distance scale as
\begin{equation}
    R = \frac{\Delta a}{l_{{\rm sh},i}+l_{{\rm sh},i+1}},
\label{eq:R}
\end{equation}
to classify the regimes. As discussed in the previous section, the vortensity perturbation $\Delta\zeta$ peaks at $1 \lesssim |r-r_{{\rm p}}|/l_{\rm sh} \lesssim 2$, so that $\pm  l_{\rm sh} $ provides with a physically-motivated scale in which vortices may be driven.

In Figure \ref{fig:Regimes_Plot} we show the final states of $37$ simulations run for $5,000$ orbits of the inner planet (see Table \ref{tab:models}) with sub-thermal masses in the range $M_{{\rm p}}/M_{\rm  th}\simeq 0.3-1$. We visually classified these simulations based on their gas morphologies after $5,000$ orbits. Similar to Figure \ref{fig:Main_Panels}, we identify systems without non-axisymmetric structures and systems with such structures lying either in between the planets or at their co-rotational regions. Thus, we propose the following classification:
\begin{description}
    \item[Regime I]The disk shows an axisymmetric ring sub-structure between orbits of the planets, forming completely separate gaps, e.g., Figure \ref{fig:Main_Panels} (b). This occurs when the planets are most widely separated with
    \begin{equation}
        R_{\rm \rm{I,II}} \equiv 3.94(M_{{\rm p}}/M_{\rm th})^{0.55}\lesssim R.
        \label{eq:R_I_II}
    \end{equation}
    \item[Regime II] The disk shows non-axisymmetric substructures between planetary orbits, with the planets sharing their partially-opened gaps, e.g., Figures \ref{fig:Main_Panels} (c) and (d). This regime takes place for intermediate planetary separations bounded by 
    \begin{equation}
       R_{\rm \rm{II,III}} \equiv 2.29(M_{{\rm p}}/M_{\rm th})^{0.34}\lesssim R\lesssim R_{\rm \rm{I,II}} .
    \end{equation}
    \item[Regime III] The disk does not show substructures between planetary orbits and they fully share their gaps. Instead, non-axisymmetric substructures do appear in their co-orbital regions around the $L_4$ or $L_5$ Lagrange points, e.g., Figure \ref{fig:Main_Panels} (e). This regime occurs for the most closely-packed systems with
        \begin{equation}
        R \lesssim R_{\rm \rm{II,III}}.
    \end{equation}
\label{itm:Regimes}
\end{description}
We remark that this proposed empirical separation has only been tested for sub-thermal planets ($M_{{\rm p}}/M_{\rm  th}\sim0.3-1$), low viscosities of $\alpha=10^{-4}$, and aspect ratios in the range $h_0\simeq0.03-0.08$.

\begin{figure}
    \centering
    \includegraphics[width=0.99\linewidth]{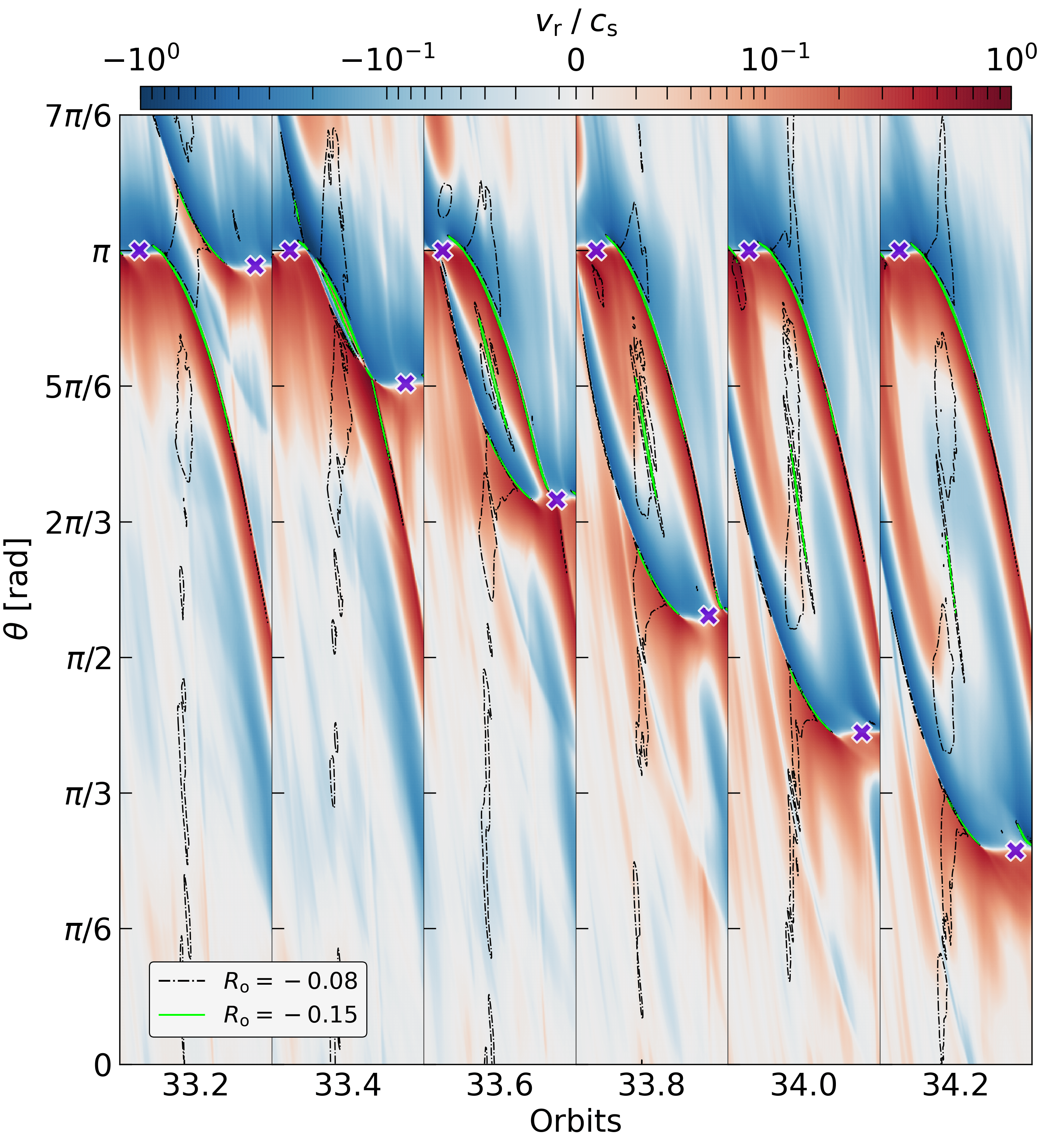}
    \caption{Evolution the fluid radial velocity $v_{\rm r}$ and overlaid black and green contours of the Rossby number $R_{\rm o}=\{-0.08, -0.15\}$ showing the early vortex development for the simulation ${\rm M1\_\Delta7.2}$. The sequence displays snapshots soon after planetary conjunction, where the interaction of the spiral density waves creates large-scale anti-cyclonic motion (clockwise) and a region with $R_{\rm o}\sim-0.15$ between the waves left behind after the wave-wave interactions. The purple crosses mark planets position and $c_{\rm s}\sim 1.53(r/\textrm{au})^{-0.357}$ km/s.}
    \label{fig:Spiral_Shock}    
\end{figure}

\begin{figure*}
\centering
\includegraphics[width=0.99\linewidth]{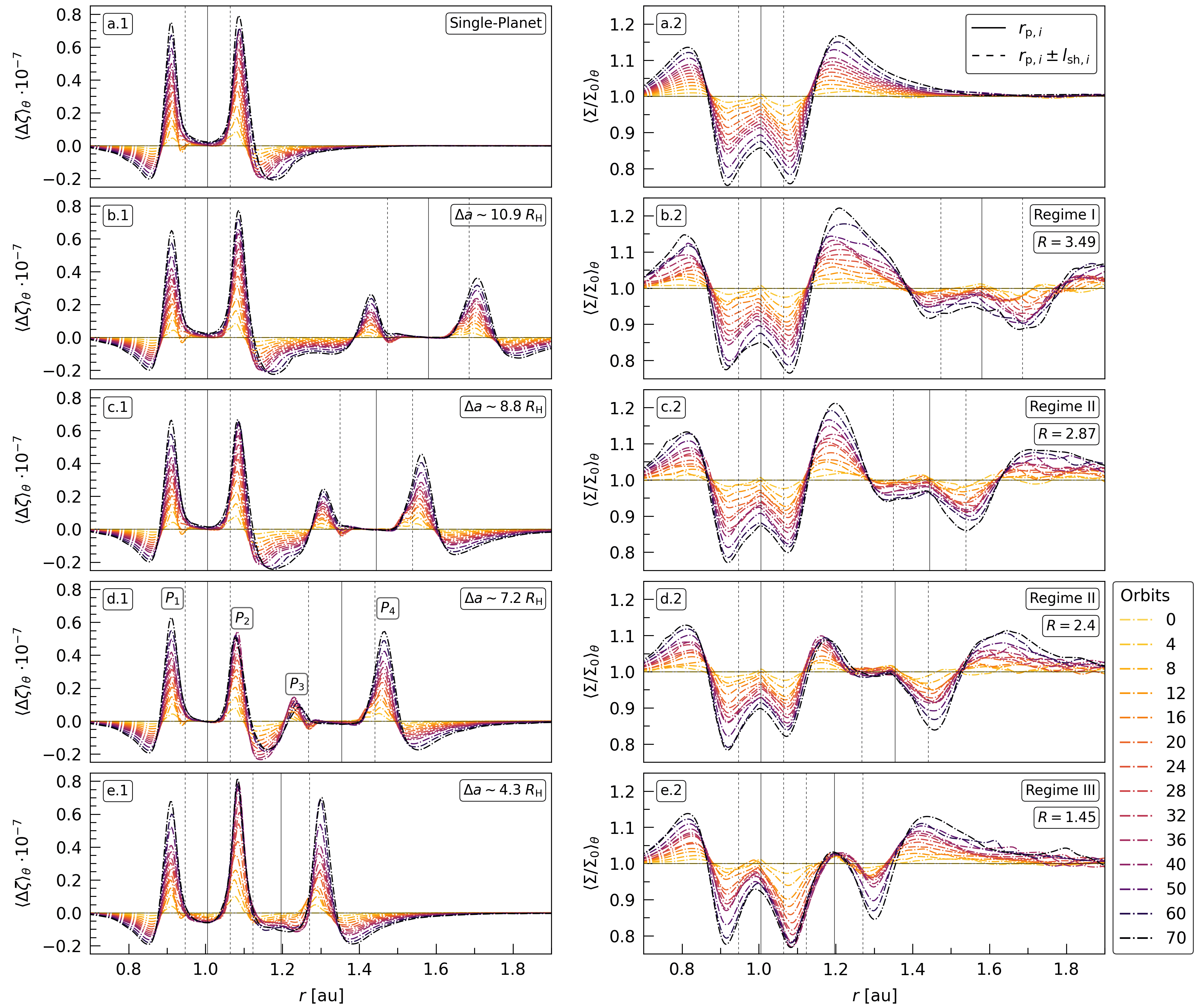}
\caption{Columns display the timeline profiles in the first $70$ orbits for vortensity perturbation $\Delta\zeta$ and surface density $\Sigma/\Sigma_0$. Solid and dashed vertical lines mark the radial position of the planet and their shock length $\pm l_{\rm sh}$ (Eq. \ref{eq:lsh}) respectively. $P_1$, $P_2$, $P_3$ and $P_4$ denote $\Delta\zeta_{{\rm peak}}$ number, where (a.1) has only $2$ peaks, while in (e.1), $P_2$ and $P_3$ are superposed. Rows correspond to different models with a single-planet (a) ${\rm M1\_\Delta0.0}$, and two-planet models (b) ${\rm M1\_\Delta10.9}$, (c) ${\rm M1\_\Delta8.8}$, (d) ${\rm M1\_\Delta7.2}$, and (e) ${\rm M1\_\Delta4.3}$.} 
\label{fig:Jump_Dens_Evolution}
\end{figure*}

\section{Partially-shared gaps (Regime II)}\label{sec:5_RegimesII}

In this section we explore the origin of the long-standing vortices in the Regime II and describe their evolution and morphology.

\subsection{The role of the secondary planet}

We begin by comparing the evolution of single-planet simulations with the same disk properties (see Table \ref{tab:models}). We run single-planet simulations with $34$, $50$, and $90$ $M_{\rm \oplus}$ and observe that they do form vortices as a result of the RWI at the gap edges. However, these merge into slightly larger vortices or a horseshoe-shaped vortex that decays away quickly spreads into a ring after $\lesssim 800$, $900$ and $1200$ orbits, respectively. These lifetimes are consistent with \citet{Hammer2021}, once the planet reaches its final mass, suggesting that the presence of a second planet is essential at sustaining and invigorating the vortex, so they live for at least $5,000$ orbits (and $10,000$ in a few extended runs).

Next, we discuss three possible roles that the second planet may play.

\subsubsection{Axisymmetric Planetary Torque}

The torque produced by the secondary planet will act in conjunction with the primary to generate a complementary gas environment to RWI. First, its density distribution opens a gap that can produce a pressure bump between planets. If the bump is sufficiently sharp, acts as a favourable condition to RWI \citep{Ono2016}. Also, the azimuthal velocity takes sub-Keplerian (super-Keplerian) values for the internal (external) disk from the planet, where a pressure bump reference frame shows a velocity field azimuthally anti-cyclonic.

In extra simulations, we applied a modified axisymmetric torque density distribution of the impulse approximation described by \citet{LinDNC1986} to emulate both conditions, where the 1D approximation is extended azimuthally to open a fully axisymmetric gap (see Appendix \ref{sec:appendix_torque} for details).

In the case of two axisymmetric gaps, there is no vortex formation anywhere in the disk. But, by embedding the low-mass planet in the internal axisymmetric gap, smaller-scale and shorter-lived vortices are formed, some of which undergo orbital migration. The end-state after 5,000 orbits shows wiggly, non-axisymmetric perturbations at the gap edges, similar to those observed in \citet[Figure 9 therein]{Hallam2017}  using  using a 1D torque\footnote{The authors implement a slightly modified torque expression from \citet{LinDNC1986} based on the impulse approximation, similar to those obtained semi-analytically by \citet{PR2012,RP2012}.}.
Therefore, even though the 1D torques (plus the small planet) may drive vortices, they may not be sufficient to sustain them for long timescales and keep them in fixed orbital distances.

Finally, the case with a planet and an axisymmetric gap, does sustain the vortex for 5,000 orbits. However, the vortex generated in the two-planet case is more compact and lower-density, with a Rossby number distribution more centrally concentrated and stronger. We suggest that such differences are created by the presence of the second planet and its launched spirals, as it directly affects the vortex region and interacts with the spiral of the primary planet as we discuss next.

\subsubsection{Spiral wave interactions} 

The spiral waves launched by the planets, interact with one another several times in different regions of the disk. 

In order to illustrate how this interaction may lead to large-scale vortices, we show in Figure \ref{fig:Spiral_Shock} the radial velocity $v_{\rm r}$ and Rossby number $R_{\rm o}$ for a sequence spanning soon after planetary conjunction. We observe that wave-wave interaction between the planets causes two effects:
\begin{enumerate}
    \item The spiral shock leaves an anticyclonic spiral-shaped remnant with $R_{\rm o} \lesssim -0.15$ (green lines). Then, it decays, elongating and dissipating azimuthally before the next conjunction.
    
    \item The propagation of spirals post-conjunction produces a favorable anti-cyclonic environment. Fundamentally, this is because the inner and outer spirals have opposite velocities (see $v_{\rm r}$ region at $\theta\sim \pi/2-5\pi/6$ after $34$ orbits), possibly assisted by the pressure bump between the planets that drives super-Keplerian (sub-Keplerian) motion in the internal (external) region.
\end{enumerate}

These two effects, namely the vortensity created from wave interactions and the anti-cyclonic environment post-conjunction, could act in concert to assist the creation and longevity of the vortices. Note these interactions occur frequently every planetary conjunction after $1/(1/P_1-1/P_2) \simeq 2.8$ yrs and only can occur in a multi planet case.

Recently, \citep{Hammer2021} has shown that the vortex migration away from the planet impacts and prolongs its evolution, predominantly caused by a weaker dissipation through shock waves. Our vortices remain long-lasting, despite of being forced to remain in between the planets.

\subsubsection{Rossby Wave Instability in a density ring}\label{sec:RWI_ring} 

As discussed in \S\ref{subsection:context}, the RWI is triggered in regions where $L_{\rm  iso}$ becomes maximum, which occurs when the vortensity perturbation $\Delta\zeta$ reaches a (negative) minimum. In Figure \ref{fig:Jump_Dens_Evolution} we display the evolution of the $\Delta\zeta$ and $\Sigma$ profiles for the first $70$ orbits to show how this criterion behaves across different regimes. 

In panel (a.1) we display the single-planet case showing the characteristic jump at $|r-r_{{\rm p}}|\gtrsim l_{\rm  sh}$, followed by a decay to negative values \citep{Dong2011b,Cimerman2021}. As such, the vortices are expected to form at $|r-r_{{\rm p}}|\sim 2 l_{\rm  sh}$, or at $0.85$ au and $1.15$ au at gap edges (panel a.2), which is indeed what we observe in our simulations (not shown). Also, we note that the slight asymmetry in the troughs and peaks around the planet is due to both the $\Delta\zeta_{{\rm peak}} \propto (M_{{\rm p}}/M_{\rm th})^{3}\propto h^{-1}$ dependency and the surface density radial profile \citep{Cimerman2021}. 

For multi-planet systems, the simulation with most widely separated planets in Regime I (panels b.1 and b.2) show that the profiles of both $\Delta\zeta$ and $\Sigma$ for the inner planet are weakly affected by the outer planet. Consistently, we checked that vortices for the inner planet develop in a similar way compared to single-planet simulations. We could regard this limit as a separable-case, similar to the simulations by \citet{Baruteau2019} for much more massive planets.

For closer-planet cases in Regime II (panels c.1 - d.3), the outer planet is able to shape the profiles of both $\Delta\zeta$ and $\Sigma$ outside the inner planets ($r\gtrsim1.2$ au). Most importantly, the region with the smallest (negative) values of $\Delta\zeta$ extend to a region around $r\sim1.2$ au, which coincides with the location of the density ring between the planets. Thus, {\it the whole ring may be subject to the RWI}. Accordingly, this is what is observed in our simulations, features that are preserved for thousands of orbits (see panels c.1 and d.2 in Figure \ref{fig:Main_Panels}).

\begin{figure}
    \centering
    \includegraphics[width=0.99\linewidth]{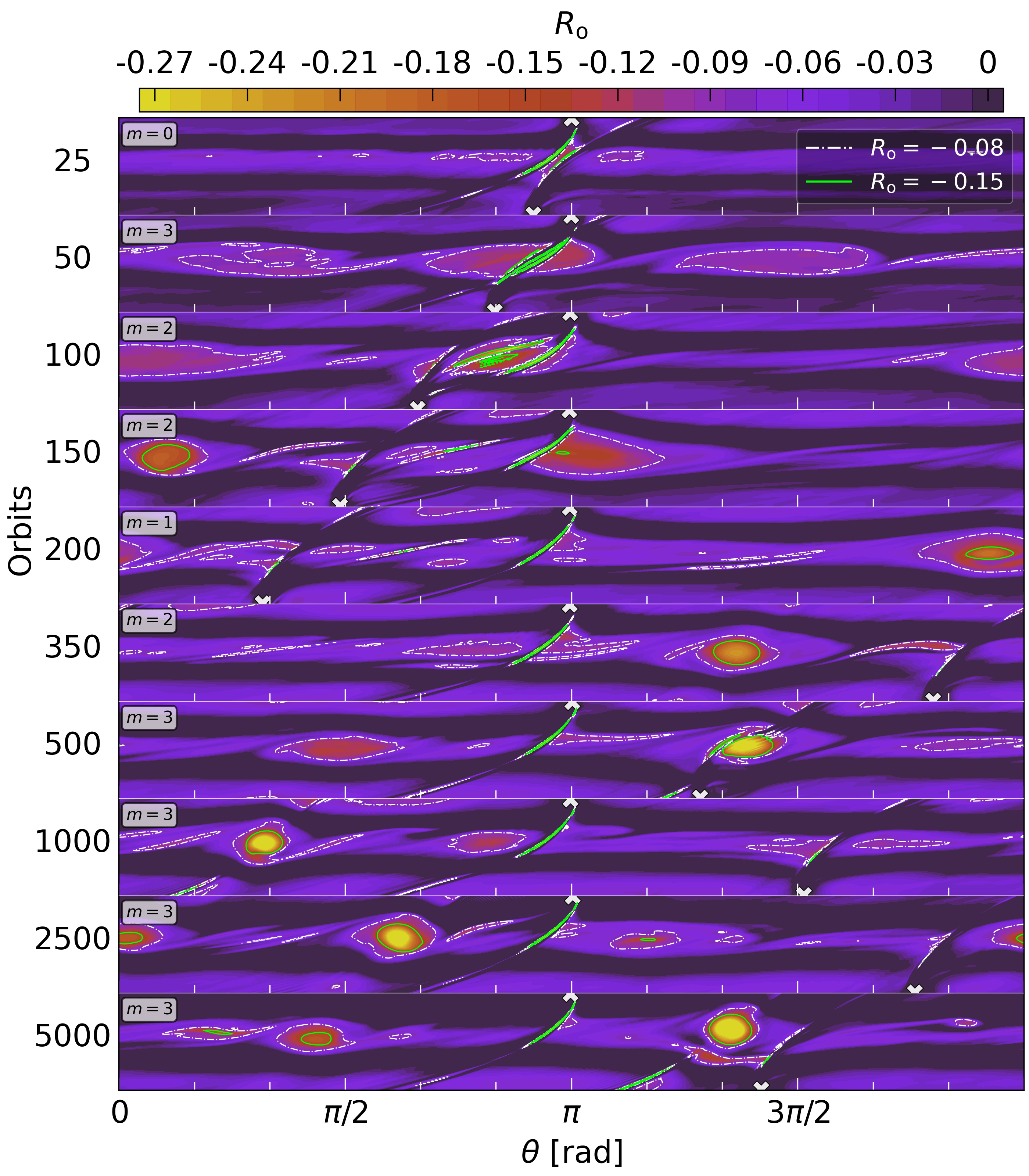}
    \caption{$R_{\rm o}$ vortices maps evolution for the radial section [$1.01$,$1.35$] au in the model ${\rm M1\_\Delta7.2}$. Panels show multiple-vortex ($m=3$) formation between 25 and 50 orbits and then its evolution in morphology and modes $m = 1$, $2$ or $3$. The overlaid white and green contours mark $R_{\rm o}=\{-0.08, -0.15\}$. White half-crosses denote planets position.}
    \label{fig:Vortex_Evolution}
\end{figure}

\subsection{Morphology of vortices}

In Figure \ref{fig:Vortex_Evolution} we show the evolution of the Rossby number for simulation ${\rm M1\_\Delta7.2}$, focusing in the radial region between the planets where the vortices develop.

We observe that the vortices start appearing after $\sim50$ orbits and with $m=3$ azimuthal symmetry and elongated. As time progresses, the number of vortices changes and becomes more compact. This general behavior is unlike previous results with a single planet where vortices tend to get more elongated and finally decay away \citep{Rometsch2021,Hammer2021}.

We observe that stronger vortices with higher values of the Rossby are more compact, as expected from the classification in \citet{Surville2015}, with elongated incompressible vortices for $R_{\rm o} > -0.15$ and compact compressible vortices for which $R_{\rm o} < -0.15$ (see the green contours in Figure \ref{fig:Vortex_Evolution}).
In order to be more quantitative, we define the aspect ratio $\chi_{R_{\rm o}} \equiv (r\Delta\theta)/\Delta r$ of the $R_{\rm o}$ contours, where $\Delta r$ and $r\Delta \theta$ are the radial and azimuthal extents measured relative the center at $r$, respectively.

We find that $\chi_{\rm -0.15} \sim 3$ when $R_{\rm o}\simeq-0.3$ at the vortex center, consistent with the value of $\sim3-4$ found by \citet{Surville2015}, who define the aspect ratio using streamlines (Figure 10 therein). Also, consistent are the density waves observed in panels (d.1) and (d.2) of Figure \ref{fig:Main_Panels}, which are expected for these compact vortices with open streamlines spiraling outward from the core \citep{bodo2005,Surville2015}.

\section{Common gaps (Regime III)}\label{sec:6_RegimesIII}

Following a similar treatment as in \S\ref{sec:RWI_ring}, in panel (e.1) of Figure \ref{fig:Jump_Dens_Evolution} we show that peaks $P_2$ and $P_3$ of the vortensity perturbation $\Delta\zeta$ are superimposed due to close proximity between the planets (compare to panel d.1). This peak falls in the region between the planets, making this region stable to RWI, which consequently either does not allow accumulations or has been quickly dissipated by wakes, unlike the Regime II case. Moreover, the negative values of $\Delta\zeta$ lie in the co-orbital regions of the planets. Thus, stable Lagrange points $L_4$ and $L_5$ may be subject to RWI to become in stronger vortices. In addition, a portion of their gas accumulations are supplied by density waves launched by the planetary companion. Both effects provide a shallower common-gap that achieves a more effective gas trapping and long-lasting agglomerations at $L_4$ and $L_5$.

\begin{figure}[htp]
    \centering
    \includegraphics[width=0.99\linewidth]{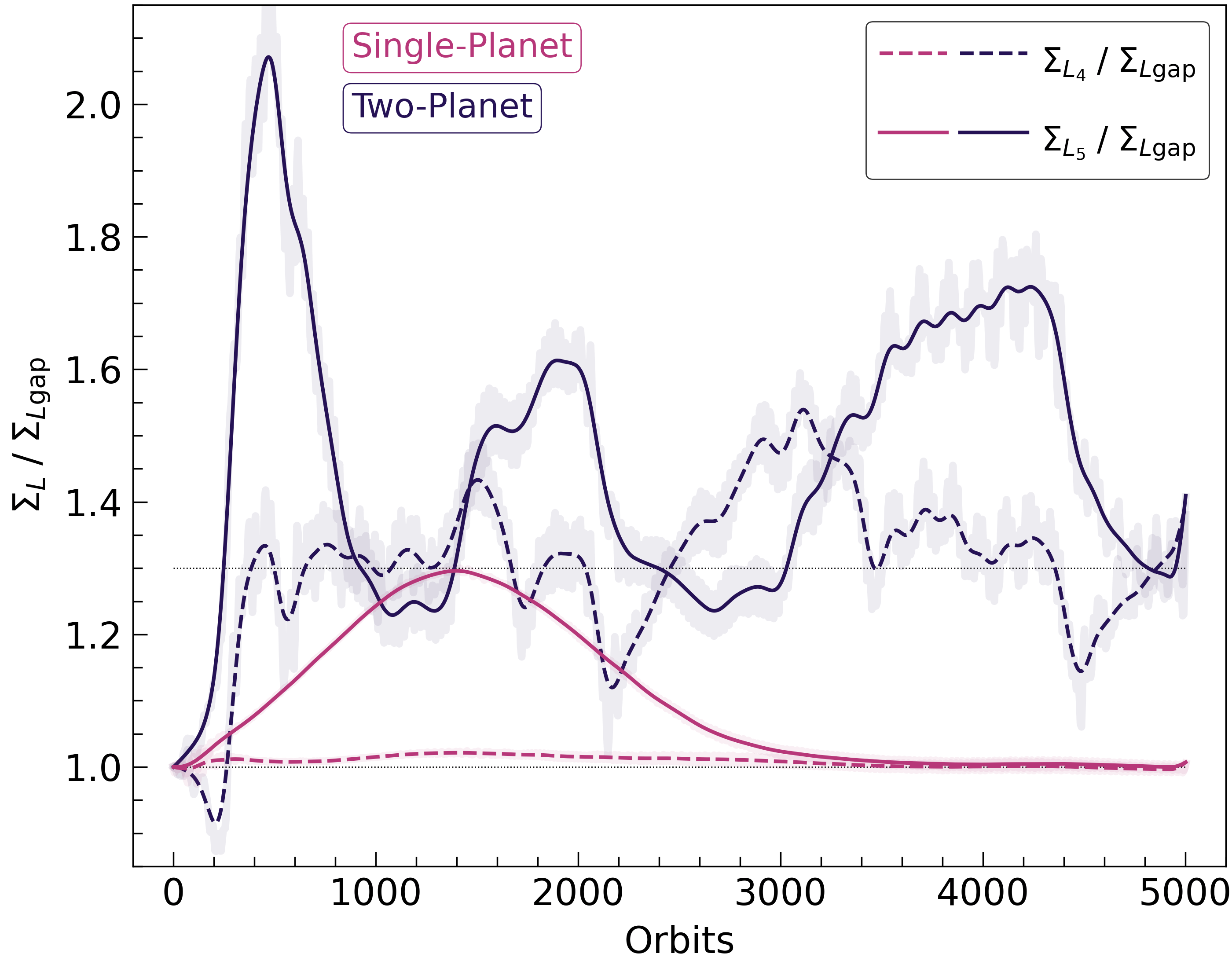}
    \caption{Evolution of the mean gas density in the vicinity of the co-orbital Lagrange points depicting the contrast between $L_4$ and $L_5$ relative to $L_{\rm gap}$. The shaded areas indicate the mean values of $\Sigma_{L_4} / \Sigma_{L{\rm gap}}$ and $\Sigma_{L_5}/\Sigma_{L{\rm gap}}$ respectively. Dashed and solid lines denote the moving average of the areas. The horizontal dotted lines at $1$ and $1.3$ denote the final values at which the models converge after $5,000$ orbits, showing that the gap in the single-planet case is emptied uniformly while two-planet systems retain significant amount gas around $L_4$ and $L_5$.}
    \label{fig:Lpoints_Evolution}
\end{figure}

\subsection{Lagrange points L4 and L5 evolution}

In order to quantify how much mass remains around the Lagrange points, we define $\Sigma_{L_4}$, $\Sigma_{L_5}$ as the mean gas surface density in the respective Lagrange points. Additionally, we denote $L_{\rm gap}$ as a local gap reference point in the co-orbital region far away from $L_4$ and $L_5$ to represent the emptier region. Radially, their widths correspond to the limits of shock surface with respect to the planet (i.e., $2l_{\rm sh}$) and azimuthally were defined by

\begin{eqnarray}
L_4:\;  \theta_{{\rm p}}+\frac{\pi}{3}-\frac{\pi}{8} &<\theta& < \theta_{{\rm p}}+\frac{\pi}{3}+\frac{\pi}{8},\\
L_5:\;  \theta_{{\rm p}}-\frac{\pi}{3}-\frac{\pi}{8} &<\theta& < \theta_{{\rm p}}-\frac{\pi}{3}+\frac{\pi}{8},\\
L_{\rm gap}:\;  \theta_{{\rm p}}+\hspace{0.20em}\pi\hspace{0.2em}-\frac{\pi}{8} &<\theta& < \theta_{{\rm p}}-\hspace{0.13em}\pi\hspace{0.2em}+\frac{\pi}{8},
\label{eq:Lpoints_Region}
\end{eqnarray} 
where $\theta_{{\rm p}}$ is the azimuthal position of the planet. 

Figure \ref{fig:Lpoints_Evolution} shows the evolution of $\Sigma_{L_4}/\Sigma_{L{\rm gap}}$ and $\Sigma_{ L_5}/\Sigma_{L{\rm gap}}$ that is expected to be larger than unity if the gas is evacuated slower in the stable Lagrange points relative to $L_{\rm gap}$. The measurements were only over the planets positioned at $r_{{\rm p}} = 1.005$ au in models ${\rm M1\_\Delta0.0}$ and ${\rm M1\_\Delta4.3}$.

The single-planet model shows an orderly evolution with $\Sigma_{L_5}/\Sigma_{L{\rm gap}}$ reaching $1.3$ after $\sim1,000$ orbits and then decaying to unity after $\sim3,000$ orbits, where the gap is azimuthally uniform. The same behavior occurs for $\Sigma_{L_4}/\Sigma_{L{\rm{gap}}}$ but with a much smaller peak amplitude, consistent with previous simulations that show more prominent asymmetries for $L_5$ \citep{Rodenkirch2021}. In contrast, the the two-planet case is far more dynamic, chaotically alternating the strength of the asymmetries between $L_4$ and $L_5$, and not settling around a single value. The final value after $5,000$ orbits of both $\Sigma_{L_4}/\Sigma_{L{\rm gap}}$ and $\Sigma_{L_5}/\Sigma_{L{\rm gap}}$ is $\sim 1.3$, similar to the peak of a single-planet case. In addition, panel (e.1) in Figure \ref{fig:Main_Panels} shows that this chaotic dynamic may switch the asymmetry between $L_4$ and $L_5$ for the outer planet, that show $L_4$ more prominent. We also tried a case with three planets, where the behavior is also quite dynamic and the asymmetries reach even larger values.

In summary, our simulations show that the lifetime and the contrast of the over-densities around $L_4$ and $L_5$ is largely increased by the presence of a nearby planetary companion.

\section{Discussion}\label{section:7_Discussion}

We have shown that sufficiently compact multi-planet systems produce long-standing non-axisymmetric substructures, under conditions where a single planet does not. Our results do not directly depend on the stellocentric distances, but on the aspect ratio $h(r) = H(r)/r$, thus allowing us to scale our predictions to more realistic, currently observable, distance scales. We discuss the conditions and possible interpretation of our findings.

\subsection{Observational signatures}

We have argued that vortices live significantly longer in compact multi-planet systems compared to single-planet cases. However, beyond the longevity and, therefore, the likelihood of detection, one would like to identify any distinguishing morphological features produced by compact multi-planet systems. We identify three.
\paragraph{Vortices within gaps}
Unlike in a single-planet system, where the RWI-driven vortices are excited at the gap edges and may migrate outwards moving away from the density gap \citep{Hammer2021}, the vortices in our Regime II remain in between the planets, well inside the gap. This occurs even if the planets undergo orbital migration.

Beyond the RWI-driven vortices, we note that crescent-shaped gas structures could be produced, in the co-orbital region of a single planet, due to the slow evacuation of gas around the stable Lagrange points. As such, the dust trapping at $L_4$ and $L_5$ could leave observable signatures for Jovian masses \citep{Montesinos2020,Rodenkirch2021}. The elongated vortices provided by Regime II show comparable morphologies to these works but with the advantage of being a less transient feature. On the other hand, our Regime III associated with the Lagrange points would produce crescent-shaped structures like in the single-planet case, but these are not centered at the gap structure. These features are reminiscent of the off-centered structure around HD 163296 \citep{Isella2018} and would require further investigation.
\paragraph{Compact vortices} 
The vortices driven by the RWI in single-planet systems are generally crescent-shaped, elongated (aspect ratios above $\sim 7$) \citep{Surville2015}, with few exceptions \citep[e.g.,][]{Les2015,Boehler2021,Hammer2021}. As we show, in our Regime II the vortices can reach aspect ratios down to $\sim 3$.

Observationally, continuum point-like emissions inside density gaps have been found such as the $6\sigma$ signal inside HD 100546 cavity \citep{Perez2020}. In a single-planet case inside a gap, the most reasonable explanation is the emission from a circumplanetary disk (CPD). In contrast, in a two-planet system, such emission may be produced by dust trapped in compact vortices. This hypothesis should be further investigated by adding dust dynamics into our gas-only simulations.
\paragraph{Shallower gaps}
The panels (a.3) and (e.3) in Figure \ref{fig:Main_Panels} show that the density gap carved by a compact planet system remains shallower than a single planet case by a factor $\sim 2$. This is due to both the remaining gas in the Lagrange points and AMF injection due to the neighboring planet may set a floor for the gap depth. The latter effect was previously discussed by \citet{DuffellDong2015} for Jovian-mass planets.

Current observations of density gaps (generally based on dust continuum) have been associated with a single planet per gap \citep{Zhang2018}.
If these observations are not able to detect such massive planets and the gaseous gaps are shallower than theoretically expected, a competitive scenario is that the gap may be carved by multiple lower-mass planets. As we describe in more detail in \S\ref{subsec:single_vs_multi}, both the depth and width of a density gap for a two-planet system cannot be reproduced by a single planet. For a given width, a single planet produces a much deeper gap.\\

We remark that these features are only based on dust-free simulations and guided by previous works (and some preliminary simulations with dust) showing correspondence between vortices and dust accumulation \citep{Brauer2008,Birnstiel2013}. In a future work, we shall consider the dust evolution in our simulations and provide more realistic and quantitative observable predictions.

\subsection{A single giant planet vs two sub-Neptunes}
\label{subsec:single_vs_multi}
As shown, $R_{\rm \rm{I,II}}$ (Eq. \ref{eq:R_I_II}) delimits when planets form separate gaps. In case to consider a Regime II, but the substructure is not prominent enough, the limit may be dropped to $\sim 0.8\times R_{\rm \rm{I,II}}$, that means, the upper limit for a system with two equal sub-thermal mass planets that share a density gap, e.g., Figure \ref{fig:Main_Panels} (c). Following \citet{Kanagawa2016} the gap width for a single planet is given by $\Delta_{{\rm gap},i} = 0.41 (m_{{\rm p},i}/M_{\rm \star})^{1/2} h^{-3/4}\alpha^{-1/4} r_{{\rm p},i}$, thus, the total gap width for a two-planet system is $\Delta_{\rm tot} = (r_{{\rm p},2}+\Delta_{\rm gap,2}/2)-(r_{{\rm p},1}-\Delta_{\rm gap,1}/2)$.

We consider the upper limit and $\alpha = 10^{-4}$ to determine the necessary single-planet mass ($M_{\rm p,single}$) positioned at $R_{\rm p,single} = (r_{{\rm p},1}+r_{{\rm p},2})/2$ to resemble $\Delta_{\rm tot}$, where using $\Delta_{\rm tot} = \Delta_{\rm gap,single}$ we roughly find that
\begin{equation}
    \frac{M_{{\rm p},{\rm single}}}{M_{\rm \oplus}} \sim 11.4 \left(\frac{h}{0.05}\right)^{1.27} \left(\frac{m_{{\rm p}}}{M_{\rm \oplus}}\right)^{0.65} \left(\frac{M_{\rm \star}}{M_{\rm \odot}}\right)^{0.35}.
    \label{eq:Mp_width}
\end{equation}

If the fiducial parameters are considered ($m_{{\rm p}} = 34 \;M_{\rm \oplus}$), Eq. \ref{eq:Mp_width} shows that a more massive planet with $M_{{\rm p}} = 117\; M_{\rm \oplus}$ is necessary. In this way, the gap resembles in width but will be affected in depth. Previous studies \citep[][hereafter K15]{Duffell2015,Kanagawa2015} have revealed that the gap depth is $\Sigma_{\rm gap}/\Sigma_0 = 1/(1+0.04K)$, where $K = q^{2} h^{-5} \alpha^{-1}$. Thus, a more massive planet will open a deeper gap, and theoretically, our single-planet gap would be $\simeq 10$ times deeper than our common gap.

Recently, the Molecules with ALMA on Planet-forming Scales (MAPS) ALMA Large Program \citep{MAPS_I} has observed the gas distributions in $5$ PPDs. A comparison with dust continuum observations shows that some deep dust continuum gaps do not have corresponding CO gaps \citep{zhang2021}, while for some gaps where there is correspondence, the gaps may be $5 - 10$ times shallower than predictions from disk-planet interactions. These predictions are based on a single planet per gap \citep{Zhang2018}, while our results show that a compact system can produce a much shallower gap by a factor of $\sim10$. Thus, placing two-Neptune instead of a Jovian may agree better with the CO MAPS results.

\subsection{Planet migration and mean motion resonances}

A well known outcome of convergent migration is the capture into mean-motion resonances (MMRs), especially into the strong first-order ones (e.g., 2:1, 3:2). Given the low surfaces densities in our simulations, the planets do not migrate, at least within $\sim10^4$ orbits. However, given our results, we may still wonder whether planets near these commensurabilities could be in the non-axisymmetric Regimes II or III.

From Figure \ref{fig:Regimes_Plot}, the condition to be in Regimes either II or III ($R<R_{\rm \rm I,II}$, Eq. \ref{eq:R_I_II}) could be roughly expressed as\footnote{We have assumed that the aspect ratio around both planets is the same, or, equivalently the flaring index $f=0$.}
\begin{equation}
\label{eq:mmr_condition}
\frac{\Delta a}{(a_{{\rm p},1}+a_{{\rm p},2})/2} \lesssim 8 h\left(\frac{M_{{\rm p}}}{M_{\rm  th}}\right)^{0.15}.
\end{equation}

Thus, at 2:1 MMR ($a_{{\rm p},2}/a_{{\rm p},1}=2^{2/3}$) and for 3:2 MMR ($a_{{\rm p},2}/a_{{\rm p},1}=1.5^{2/3}$), the condition for the aspect ratio and planet mass from Eq. \ref{eq:mmr_condition} reads
\begin{eqnarray}
0.057&\lesssim& h\left(\frac{M_{{\rm p}}}{M_{\rm  th}}\right)^{0.15} \quad\quad \textrm{2:1 MMR},\\
0.033&\lesssim& h\left(\frac{M_{{\rm p}}}{M_{\rm  th}}\right)^{0.15} \quad\quad \textrm{3:2 MMR}.
\end{eqnarray}

We recall that these estimates have been tested for a limited range of parameters and sub-thermal-mass planets ($M_{{\rm p}}/M_{\rm  th}\sim0.3-1$), or $h\simeq 0.03-0.06$ for near Neptune-mass planets. Thus, vortices between the planets may be common in systems captured into 3:2 MMRs, but less commonly expected in systems around the 2:1 MMRs. As a proof of concept, we ran the ${\rm M1\_\Delta8.8}$ model increasing the surface density by a factor of 400 to $\Sigma_0 = 680\;(r/r_0)^{-1.5} \;\rm gr/cm^2$. Its planet separation start in $(a_{{\rm p},2}/a_{{\rm p},1})^{3/2} = 1.72$, wider than 3:2 MMR. We check that a Regime II$-$like vortex is formed and migrate with the planets. Consistent with our estimates, the recent simulations by \citet{Cui2021} including migration and crossing of the 3:2 MMR, do not show evidence vortices at $\sim8000$ orbits as the aspect ratio is too low ($h=0.02$, Figure 2 therein). A more in-depth study of the dynamics including migration may be required to establish whether vortices within gaps are a consistent signature of convergent planet migration. 

\subsection{Outer Solar System: Neptune and Uranus}

An interesting question is whether the planets in our Solar System may be detectable if placed around another star. Given our proposal that Neptune-mass planets may exhibit readily observable vortices, we assess whether the  Uranus and Neptune pair lie in Regimes II/III,  especially interesting as the outer-most planets are most easily resolvable given our current spatial resolution constraints.

For the current orbital separations of Uranus and Neptune with $\Delta a/a_{{\rm p},1}= 0.56$, we find that a Regime II would require an aspect ratio $h \gtrsim 0.078$, thus suggesting that the Regime I is more likely. However, in the disk-embedded phases, their relative separations were likely smaller given Neptune's outward planetesimal-driven migration from $a_{\rm N}\lesssim 25$ au ($\Delta a/a_{{\rm p},1}\lesssim 0.3$) \citep{Malhotra1993} or an earlier period of dynamical upheaval\footnote{The so-called Nice model suggests that Uranus and Neptune started more compact at $\sim 17 $ au and $\sim 12 $ au ($\Delta a/a_{{\rm p},1}\lesssim 0.4$) and crossed their orbits \citep{Tsiganis2005}.}. For reference, if $\Delta a/a_{{\rm p},1}\lesssim 0.4$ we find that the Regime II is reached for $h\gtrsim 0.049$. We have confirmed these estimates by running a simulation with Uranus and Neptune separated by $\Delta a/a_{{\rm p},1}=0.4$ using our fiducial parameters ($h=0.051$) and find that an elongated vortex forms in between the planets.

\subsection{Viscosity and Thermodynamic Effects}

\citet{Miranda2020} explored the role of viscosity in density wave propagation, showed that viscous damping is subdominant compared with nonlinear effects. It becomes competitive just for $\alpha \gtrsim 10^{-2}$. Also, studies by analyzing spectral line broadening \citep{Flaherty2015,Flaherty2018} have shown low-turbulence values ($\alpha \lesssim 10^{-3}$) in discs. However, non-zero viscosity play a role in gap depth through $K \propto \alpha^{-1}$ \citep{Duffell2015,Duffell2020,Kanagawa2015,Kanagawa2017,Dempsey2020} as $\Sigma_{\rm gap}/\Sigma_0 = 1/(1+0.04K)$. Therefore, $\alpha$ is not essential to classify regimens but so important to contrast the substructures with respect to the gap,  obtaining more prominent agglomerations for larger $K$. In addition, \citet{Hammer2021} shows that vortices produced at the edges of the gap are long-lived when are triggered by lower $m_{{\rm p}}$, smaller $h$, and lower $\alpha$ values.

Radiative damping can play an important role in the wave dissipation \citep{Miranda2020} and for cooling time in the range $t_{\rm cool}\Omega_{\rm K}\in [0.1,1]$, the planetary spiral waves become weaker and the AMF dissipates faster compared to the the spirals in an isothermal disk \citep{Zhang&Zhu2020}. Accordingly, the vortices' lifetime may be affected by considering a finite cooling time as shown by \citet{Rometsch2021}, where the shortest lifetimes occur for $t_{\rm cool}\Omega_{\rm K} \sim 1$ \citep{Fung2021}. A similar effect may be expected in our simulations, but possibly less important as the vortices' lifetimes in simulations with multiple planets depend on more factors than just the wave damping by shocks. We defer this study for future work.

\begin{deluxetable*}{l|lcl|ccr|ccc|c}
\tablecolumns{11}
\tablecaption{Parameters of the models}
\tablehead{
 \colhead{Model}    & 
 \colhead{$h_0$}    & 
 \colhead{$\alpha$} & 
 \colhead{$n_{\rm r} \times n_{\rm \theta}$}    &
 \colhead{$m_{\rm  p}$} &
 \colhead{$r_{{\rm p}}$}   & 
 \colhead{$\Delta$} &
 \colhead{$l_{\rm sh}$}  & 
 \colhead{$K$}            & 
 \colhead{$R$}            & 
 \colhead{Regime}       \\
 \nocolhead{} &
 \nocolhead{} &
 \nocolhead{} &
 \colhead{[cells $\times$ cells]} &
 \colhead{[$M_{\rm {\oplus}}$]} &
 \colhead{[au]} &
 \colhead{[$R_{\rm H}$]} &
 \colhead{[au]} &
 \colhead{(\citetalias{Kanagawa2015})} &
 \colhead{(Eq. \ref{eq:R})} &
 \colhead{(\S \ref{subsec:Regimes})}
 }
\startdata
M1\_$\Delta$0.0  & 0.0516 & $10^{-4}$ & 768 $\times$ 1280 & 34     & 1.005        & |    & 0.059        & 284      & |    & |\\
M1\_$\Delta$4.3  & 0.0516 & $10^{-4}$ & 768 $\times$ 1280 & 34, 34 & 1.005, 1.197 & 4.26 & 0.059, 0.074 & 283, 250 & 1.45 & III \\
M1\_$\Delta$5.6 & 0.0516 & $10^{-4}$ & 512 $\times$ 768 & 34, 34 & 1.005, 1.264 & 5.59 & 0.059, 0.079 & 283, 240 & 1.88 & III \\
M1\_$\Delta$6.0 & 0.0516 & $10^{-4}$ & 512 $\times$ 768 & 34, 34 & 1.005, 1.286 & 6.01 & 0.059, 0.081 & 283, 237 & 2.02 & III \\
M1\_$\Delta$6.4 & 0.0516 & $10^{-4}$ & 512 $\times$ 768 & 34, 34 & 1.005, 1.309 & 6.43 & 0.059, 0.083 & 283, 234 & 2.15 & II \\
M1\_$\Delta$7.2  & 0.0516 & $10^{-4}$ & 768 $\times$ 1280 & 34, 34 & 1.005, 1.354 & 7.24 & 0.059, 0.087 & 283, 229 & 2.40 & II \\
M1\_$\Delta$8.0 & 0.0516 & $10^{-4}$ & 512 $\times$ 768 & 34, 34 & 1.005, 1.399 & 8.03 & 0.059, 0.09 & 283, 223 & 2.65 & II \\
M1\_$\Delta$8.8  & 0.0516 & $10^{-4}$ & 768 $\times$ 1280 & 34, 34 & 1.005, 1.444 & 8.78 & 0.059, 0.094 & 283, 218 & 2.87 & II \\
M1\_$\Delta$9.5 & 0.0516 & $10^{-4}$ & 512 $\times$ 768 & 34, 34 & 1.005, 1.489 & 9.51 & 0.059, 0.098 & 283, 214 & 3.09 & II \\
M1\_$\Delta$9.9 & 0.0516 & $10^{-4}$ & 512 $\times$ 768 & 34, 34 & 1.005, 1.512 & 9.86 & 0.059, 0.100 & 283, 211 & 3.19 & II \\
M1\_$\Delta$10.2 & 0.0516 & $10^{-4}$ & 512 $\times$ 768 & 34, 34 & 1.005, 1.535 & 10.22 & 0.059, 0.102 & 283, 209 & 3.30 & II \\
M1\_$\Delta$10.9  & 0.0516 & $10^{-4}$ & 768 $\times$ 1280 & 34, 34 & 1.005, 1.580 & 10.89 & 0.059, 0.106 & 283, 205 & 3.49 & I \\
M2\_$\Delta$6.0 & 0.0516 & $10^{-4}$ & 512 $\times$ 768 & 30, 30 & 1.005, 1.272 & 5.99 & 0.062, 0.084 & 220, 186 & 1.83 & III \\
M2\_$\Delta$6.9 & 0.0516 & $10^{-4}$ & 512 $\times$ 768 & 30, 30 & 1.005, 1.317 & 6.87 & 0.062, 0.088 & 220, 181 & 2.09 & II \\
M2\_$\Delta$8.5 & 0.0516 & $10^{-4}$ & 512 $\times$ 768 & 30, 30 & 1.005, 1.408 & 8.52 & 0.062, 0.096 & 220, 173 & 2.56 & II \\
M2\_$\Delta$9.7 & 0.0516 & $10^{-4}$ & 512 $\times$ 768 & 30, 30 & 1.005, 1.476 & 9.69 & 0.062, 0.102 & 220, 167 & 2.88 & II \\
M2\_$\Delta$11.2 & 0.0516 & $10^{-4}$ & 512 $\times$ 768 & 30, 30 & 1.005, 1.566 & 11.16 & 0.062, 0.110 & 220, 160 & 3.26 & I \\
M3\_$\Delta$7.1 & 0.0516 & $10^{-4}$ & 512 $\times$ 768 & 15, 15 & 1.005, 1.255 & 7.12 & 0.081, 0.109 & 55, 46 & 1.31 & III \\
M3\_$\Delta$10.4 & 0.0516 & $10^{-4}$ & 512 $\times$ 768 & 15, 15 & 1.005, 1.394 & 10.44 & 0.081, 0.125 & 55, 43 & 1.89 & II \\
M3\_$\Delta$12.0 & 0.0516 & $10^{-4}$ & 512 $\times$ 768 & 15, 15 & 1.005, 1.465 & 11.99 & 0.081, 0.133 & 55, 42 & 2.14 & I \\
M4\_$\Delta$3.1 & 0.06 & $10^{-4}$ & 640 $\times$ 1024 & 30, 30 & 1.005, 1.135 & 3.10 & 0.086, 0.101 & 103, 95$\;$ & 0.70 & III \\
M4\_$\Delta$4.5 & 0.06 & $10^{-4}$ & 640 $\times$ 1024 & 30, 30 & 1.005, 1.199 & 4.51 & 0.086, 0.108 & 103, 91$\;$ & 1.00 & III \\
M4\_$\Delta$5.8 & 0.06 & $10^{-4}$ & 640 $\times$ 1024 & 30, 30 & 1.005, 1.264 & 5.84 & 0.086, 0.116 & 103, 88$\;$ & 1.29 & III \\
M4\_$\Delta$7.1 & 0.06 & $10^{-4}$ & 640 $\times$ 1024 & 30, 30 & 1.005, 1.329 & 7.09 & 0.086, 0.124 & 103, 85$\;$ & 1.55 & III \\
M4\_$\Delta$9.0 & 0.06 & $10^{-4}$ & 640 $\times$ 1024 & 30, 30 & 1.005, 1.437 & 9.04 & 0.086, 0.137 & 103, 80$\;$ & 1.94 & II \\
M4\_$\Delta$10.9 & 0.06 & $10^{-4}$ & 512 $\times$ 768 & 30, 30 & 1.005, 1.549 & 10.88 & 0.086, 0.151 & 103, 76$\;$ & 2.29 & II \\
M4\_$\Delta$12.7 & 0.06 & $10^{-4}$ & 512 $\times$ 768 & 30, 30 & 1.005, 1.667 & 12.65 & 0.086, 0.167 & 103, 72$\;$ & 2.62 & I \\
M5\_$\Delta$4.5 & 0.04 & $10^{-4}$ & 640 $\times$ 1024 & 10, 10 & 1.005, 1.135 & 4.46 & 0.055, 0.064 & 87, 80 & 1.09 & III \\
M5\_$\Delta$6.5 & 0.04 & $10^{-4}$ & 640 $\times$ 1024 & 10, 10 & 1.005, 1.199 & 6.50 & 0.055, 0.069 & 87, 77 & 1.58 & III \\
M5\_$\Delta$7.2 & 0.04 & $10^{-4}$ & 640 $\times$ 1024 & 10, 10 & 1.005, 1.221 & 7.15 & 0.055, 0.070 & 87, 76 & 1.73 & III \\
M5\_$\Delta$8.4 & 0.04 & $10^{-4}$ & 640 $\times$ 1024 & 10, 10 & 1.005, 1.264 & 8.42 & 0.055, 0.074 & 87, 74 & 2.02 & II \\
M5\_$\Delta$10.2 & 0.04 & $10^{-4}$ & 640 $\times$ 1024 & 10, 10 & 1.005, 1.329 & 10.23 & 0.055, 0.079 & 87, 71 & 2.43 & II \\
M5\_$\Delta$11.8 & 0.04 & $10^{-4}$ & 512 $\times$ 768 & 10, 10 & 1.005, 1.388 & 11.78 & 0.055, 0.083 & 87, 69 & 2.77 & I \\
M6\_$\Delta$4.5 & 0.0316 & $10^{-4}$ & 640 $\times$ 1024 & 10, 10 & 1.005, 1.135 & 4.46 & 0.033, 0.038 & 283, 260 & 1.83 & III \\
M6\_$\Delta$6.5 & 0.0316 & $10^{-4}$ & 640 $\times$ 1024 & 10, 10 & 1.005, 1.199 & 6.50 & 0.033, 0.041 & 283, 249 & 2.64 & II \\
M6\_$\Delta$8.4 & 0.0316 & $10^{-4}$ & 640 $\times$ 1024 & 10, 10 & 1.005, 1.264 & 8.42 & 0.033, 0.044 & 283, 240 & 3.39 & II \\
M6\_$\Delta$9.6 & 0.0316 & $10^{-4}$ & 640 $\times$ 1024 & 10, 10 & 1.005, 1.308 & 9.64 & 0.033, 0.046 & 283, 235 & 3.85 & I \\
M7\_$\Delta$8.4 & 0.08 & $10^{-4}$ & 512 $\times$ 768 & 100, 100 & 1.005, 1.655 & 8.35 & 0.100, 0.192 & 273, 191 & 2.23 & II \\
M8\_$\Delta$0.0 & 0.0516 & $10^{-4}$ & 768 $\times$ 1280 & 50     & 1.005        & |    & 0.050        & 612      & |    & |\\
M9\_$\Delta$0.0 & 0.0516 & $10^{-4}$ & 768 $\times$ 1280 & 90     & 1.005        & |    & 0.039        & 1984      & |    & |\\
MSS\_$\Delta$NU & 0.0516 & $10^{-4}$ & 768 $\times$ 1280 & 15, 17 & 12.56, 17.59 & 10.52 & 1.029, 1.495 & 51, 57$\;$ & 1.99 & II \\
P\_2Dg & 0.0516 & $10^{-4}$ & 512 $\times$ 768 & 34, 34\tablenotemark{\scriptsize l} & 1.005, 1.444 & 8.78 & 0.059 & 283 & | & II \\
2DgsP\_2Dg & 0.0516 & $10^{-4}$ & 512 $\times$ 768 & 34\tablenotemark{\scriptsize l},1, 34\tablenotemark{\scriptsize l} & 1.0,1.0, 1.44 & 8.78 & 0.240 & 0.25 & | & I \\
2Dg\_2Dg & 0.0516 & $10^{-4}$ & 512 $\times$ 768 & 34\tablenotemark{\scriptsize l}, 34\tablenotemark{\scriptsize l} & 1.005, 1.444 & 8.78 & | & | & | & I \\
\enddata
\tablenotetext{\tiny l}{Corresponding planetary masses used in \citet{LinDNC1986} modified formula, see Apendix \ref{sec:appendix_torque}.}
\label{tab:models}
\end{deluxetable*}

\section{Conclusions}
\label{section:8_Conclutions}

We performed 2D hydrodynamic simulations of compact systems in the sub-thermal mass regime to study the disk's surface density evolution. 

We identify two regimes that create  non-axisymmetric structures, depending on the interplanetary separation $\Delta a$ in units of $H_{{\rm p}}$ (the average disk scale height around the planets):
\begin{itemize}
    \item For $\Delta a/H_{{\rm p}}\sim 4.6-8.0\; $, a density ring that separates the planets becomes unstable to the RWI forming vortices with a wide range of aspects ratios down to $\sim 3$. 
    
    \item For the most compact systems with $\Delta a/H_{{\rm p}}\lesssim 4.6$, the planets fully share their density gaps and present long-lived elongated vortices around the stable co-rotational Lagrange points.
\end{itemize}
The vortices in both cases do not show any signs of dissipating after 5,000 - 10,000 orbits and we suggest that their longevity is promoted by their frequent interaction of spiral density waves launched by the neighboring planets.

Overall, our results show that long-lived non-axisymmetric structures {\it inside} density gaps are prevalent in compact systems with Neptune-mass planets, in contrast to single planets that create vortices at the gap edges. This distinctive feature, along with the shallower density gaps, may be smoking guns of compact multi-planet systems, possibly trapped into mean-motion resonances.\\

\begin{acknowledgements}
The authors would like to thank Andrew Youdin, Chris Matzner,  Diego Muñoz, Pablo Ben\'itez-Llambay and Ximena Ramos for helpful discussions that improved the quality of this paper and Juan Veliz for his support with the cluster logistics. 
J.G. acknowledge support by ANID, -- Millennium Science Initiative Program -- NCN19\_171 and FONDECYT Regular grant 1210425.
C.P. acknowledges support from ANIDMillennium Science Initiative-ICN12\_009, CATA-Basal AFB-170002, ANID BASAL project FB210003, FONDECYT Regular grant 1210425 and ANID + REC Convocatoria Nacional subvencion a la instalacion en la Academia convocatoria 2020 PAI77200076.
The Geryon cluster at the Centro de Astro-Ingenieria UC was extensively used for the calculations performed in this paper. BASAL CATA PFB-06, the Anillo ACT-86, FONDEQUIP AIC-57, and QUIMAL 130008 provided funding for several improvements to the Geryon cluster.
\end{acknowledgements}

\software{\textsc{Fargo3D} \citep{Benitez2016}, \textsc{Numpy} \citep{vanderWalt2011}, \textsc{Matplotlib} \citep{Hunter2007}.}

\appendix

\section{Simulations with a 1D axisymmetric planetary torque} \label{sec:appendix_torque}

In order to study the influence of planetary torque on the vortex generation and lifetime, we perform extra simulations that include the torque density distribution of the impulse approximation described by \citet[Eq. 14 therein]{LinDNC1986}. A multiplicative factor of 0.3 is used to match the depth of the gaps to those in our simulations (otherwise the gaps are too deep). The 1D approximation is extended over the azimuth to open a fully axisymmetric gap, we will refer to this as 2Dgap.

Three configurations have been tested to compare the planet$-$planet case (model ${\rm M1\_\Delta8.8}$), namely: the planet$-$2Dgap, 2Dgap$-$2Dgap and (2Dgap$+$small planet)$-$2Dgap. In the latter case we embed a low-mass planet of mass $1\;M_{\oplus}$ at the center of the inner gap open by the axisymmetric torque to drive radial perturbations from spiral density waves. 

\begin{figure}
    \centering
    \includegraphics[width=0.5\linewidth]{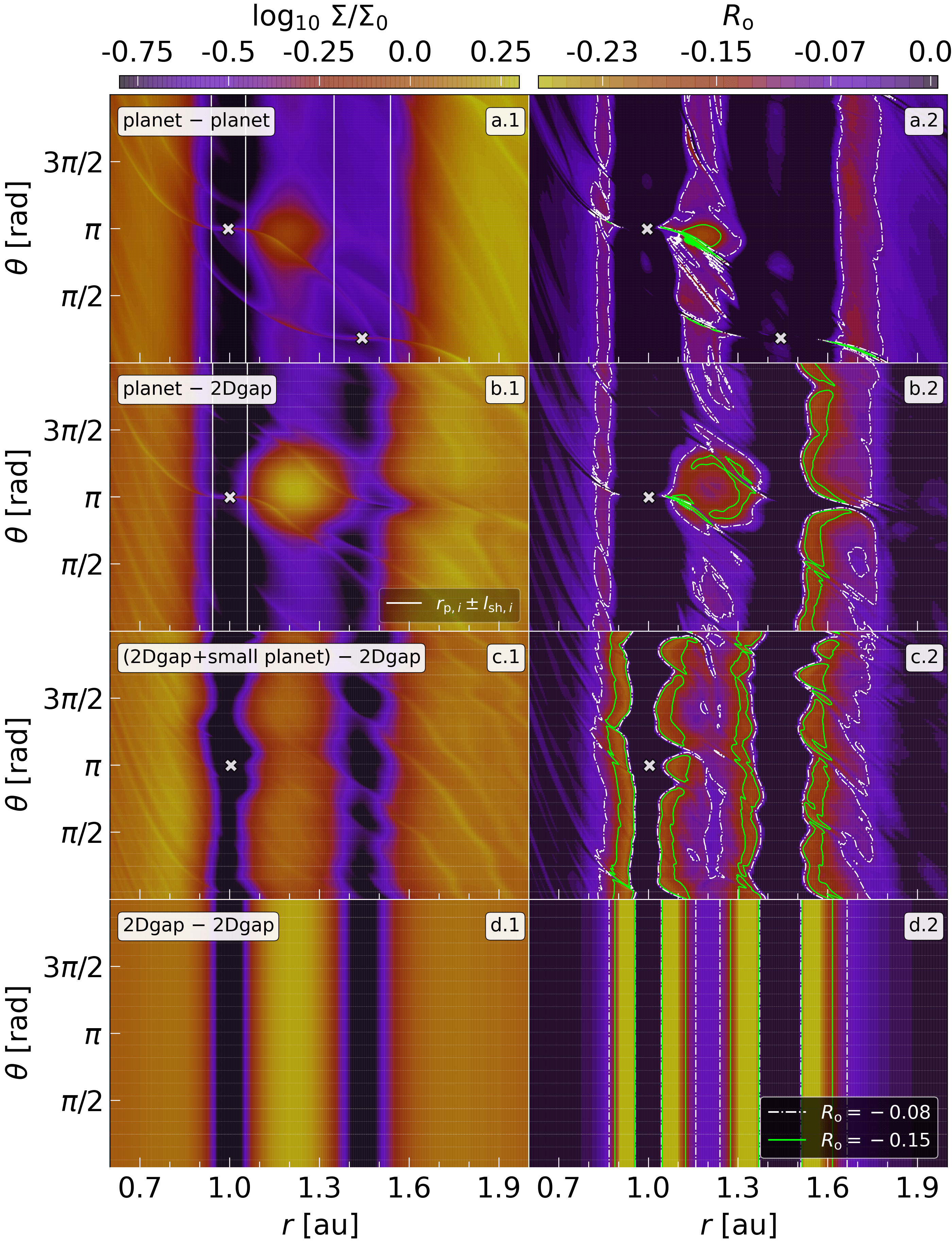}
    \caption{Columns show snapshots after 5000 orbits at 1 au. Left: Maps $\Sigma/\Sigma_0$ in log-scale. The white vertical lines indicate the shock length $\pm l_{\rm sh}$ (Eq. \ref{eq:lsh}) from each planet, where the crosses denotes their position. Right: Contours of the Rossby number $R_{\rm o}$, constrained to negative values displaying anticyclonic rotation. White and green, contours denote $R_{\rm o} = {-0.08,-0.15}$. Rows correspond to configurations (a) planet$-$planet, (b) planet$-$2Dgap, (c) (2Dgap+small planet)$-$2Dgap and (d) 2Dgap$-$2Dgap, described in Appendix \ref{sec:appendix_torque}, where their models are ${\rm M1\_\Delta8.8}$, ${\rm P\_2Dg}$, ${\rm 2DgsP\_2Dg}$ and ${\rm 2Dg\_2Dg}$ respectively, see Table \ref{tab:models} for details.}
    \label{fig:Torque}
\end{figure}

The Figure \ref{fig:Torque} shows all the cases after $5,000$ orbits and reveals that in a 2Dgap$-$2Dgap case, there is no vortex formation anywhere in the disk. The planet$-$2Dgap case resembles the behavior of the planet$-$planet case, showing a vortex with higher density values and larger radial extension. However, its Rossby number distribution is non-centrally concentrated and weaker in the center. Finally, the (2Dgap$+$small planet)$-$2Dgap case shows a turbulent environment, but no vortex has survived. In the early stages, this configuration forms two main vortices around $1.2$ au. While one of them decays after $1,200$ orbits, the other migrates inward across the inner gap ($400$ orbits) to decay in the inner disk after $1,000$ orbits. We have also appreciated this vortex migration effect for a planet$-$2Dgap case with a more compact configuration (model ${\rm M1\_\Delta7.2}$).

\bibliography{References}{}

\begin{thebibliography}{}
\expandafter\ifx\csname natexlab\endcsname\relax\def\natexlab#1{#1}\fi
\providecommand{\url}[1]{\href{#1}{#1}}
\providecommand{\dodoi}[1]{doi:~\href{http://doi.org/#1}{\nolinkurl{#1}}}
\providecommand{\doeprint}[1]{\href{http://ascl.net/#1}{\nolinkurl{http://ascl.net/#1}}}
\providecommand{\doarXiv}[1]{\href{https://arxiv.org/abs/#1}{\nolinkurl{https://arxiv.org/abs/#1}}}

\bibitem[{{ALMA Partnership} {et~al.}(2015){ALMA Partnership}, {Brogan},
  {P{\'e}rez}, {Hunter}, {Dent}, {Hales}, {Hills}, {Corder}, {Fomalont},
  {Vlahakis}, {Asaki}, {Barkats}, {Hirota}, {Hodge}, {Impellizzeri}, {Kneissl},
  {Liuzzo}, {Lucas}, {Marcelino}, {Matsushita}, {Nakanishi}, {Phillips},
  {Richards}, {Toledo}, {Aladro}, {Broguiere}, {Cortes}, {Cortes}, {Espada},
  {Galarza}, {Garcia-Appadoo}, {Guzman-Ramirez}, {Humphreys}, {Jung}, {Kameno},
  {Laing}, {Leon}, {Marconi}, {Mignano}, {Nikolic}, {Nyman}, {Radiszcz},
  {Remijan}, {Rod{\'o}n}, {Sawada}, {Takahashi}, {Tilanus}, {Vila Vilaro},
  {Watson}, {Wiklind}, {Akiyama}, {Chapillon}, {de Gregorio-Monsalvo}, {Di
  Francesco}, {Gueth}, {Kawamura}, {Lee}, {Nguyen Luong}, {Mangum}, {Pietu},
  {Sanhueza}, {Saigo}, {Takakuwa}, {Ubach}, {van Kempen}, {Wootten},
  {Castro-Carrizo}, {Francke}, {Gallardo}, {Garcia}, {Gonzalez}, {Hill},
  {Kaminski}, {Kurono}, {Liu}, {Lopez}, {Morales}, {Plarre}, {Schieven},
  {Testi}, {Videla}, {Villard}, {Andreani}, {Hibbard}, \&
  {Tatematsu}}]{ALMA2015}
{ALMA Partnership}, {Brogan}, C.~L., {P{\'e}rez}, L.~M., {et~al.} 2015, \apjl,
  808, L3, \dodoi{10.1088/2041-8205/808/1/L3}

\bibitem[{{Andrews}(2020)}]{Andrews2020}
{Andrews}, S.~M. 2020, \araa, 58, 483,
  \dodoi{10.1146/annurev-astro-031220-010302}

\bibitem[{{Bae} {et~al.}(2018){Bae}, {Pinilla}, \& {Birnstiel}}]{Bae2018}
{Bae}, J., {Pinilla}, P., \& {Birnstiel}, T. 2018, \apjl, 864, L26,
  \dodoi{10.3847/2041-8213/aadd51}

\bibitem[{{Bae} {et~al.}(2019){Bae}, {Zhu}, {Baruteau}, {Benisty}, {Dullemond},
  {Facchini}, {Isella}, {Keppler}, {P{\'e}rez}, \& {Teague}}]{Bae2019}
{Bae}, J., {Zhu}, Z., {Baruteau}, C., {et~al.} 2019, \apjl, 884, L41,
  \dodoi{10.3847/2041-8213/ab46b0}

\bibitem[{{Baruteau} {et~al.}(2019){Baruteau}, {Barraza}, {P{\'e}rez},
  {Casassus}, {Dong}, {Lyra}, {Marino}, {Christiaens}, {Zhu}, {Carmona},
  {Debras}, \& {Alarcon}}]{Baruteau2019}
{Baruteau}, C., {Barraza}, M., {P{\'e}rez}, S., {et~al.} 2019, \mnras, 486,
  304, \dodoi{10.1093/mnras/stz802}

\bibitem[{{Ben{\'\i}tez-Llambay} {et~al.}(2019){Ben{\'\i}tez-Llambay}, {Krapp},
  \& {Pessah}}]{Benitez2019}
{Ben{\'\i}tez-Llambay}, P., {Krapp}, L., \& {Pessah}, M.~E. 2019, \apjs, 241,
  25, \dodoi{10.3847/1538-4365/ab0a0e}

\bibitem[{{Ben{\'\i}tez-Llambay} \& {Masset}(2016)}]{Benitez2016}
{Ben{\'\i}tez-Llambay}, P., \& {Masset}, F.~S. 2016, \apjs, 223, 11,
  \dodoi{10.3847/0067-0049/223/1/11}

\bibitem[{{Birnstiel} {et~al.}(2013){Birnstiel}, {Dullemond}, \&
  {Pinilla}}]{Birnstiel2013}
{Birnstiel}, T., {Dullemond}, C.~P., \& {Pinilla}, P. 2013, \aap, 550, L8,
  \dodoi{10.1051/0004-6361/201220847}

\bibitem[{{Bodo} {et~al.}(2005){Bodo}, {Chagelishvili}, {Murante}, {Tevzadze},
  {Rossi}, \& {Ferrari}}]{bodo2005}
{Bodo}, G., {Chagelishvili}, G., {Murante}, G., {et~al.} 2005, \aap, 437, 9,
  \dodoi{10.1051/0004-6361:20041046}

\bibitem[{{Boehler} {et~al.}(2021){Boehler}, {M{\'e}nard}, {Robert}, {Isella},
  {Pinte}, {Gonzalez}, {van der Plas}, {Weaver}, {Teague}, {Garg}, \&
  {M{\'e}heut}}]{Boehler2021}
{Boehler}, Y., {M{\'e}nard}, F., {Robert}, C.~M.~T., {et~al.} 2021, \aap, 650,
  A59, \dodoi{10.1051/0004-6361/202040089}

\bibitem[{Bowler \& Nielsen(2018)}]{Bowler2018}
Bowler, B.~P., \& Nielsen, E.~L. 2018, Occurrence Rates from Direct Imaging
  Surveys (Cham: Springer International Publishing), 1967--1983,
  \dodoi{"10.1007/978-3-319-55333-7_155"}

\bibitem[{{Brauer} {et~al.}(2008){Brauer}, {Dullemond}, \&
  {Henning}}]{Brauer2008}
{Brauer}, F., {Dullemond}, C.~P., \& {Henning}, T. 2008, \aap, 480, 859,
  \dodoi{10.1051/0004-6361:20077759}

\bibitem[{{Bro{\v{z}}} {et~al.}(2018){Bro{\v{z}}}, {Chrenko}, {Nesvorn{\'y}},
  \& {Lambrechts}}]{Broz2018}
{Bro{\v{z}}}, M., {Chrenko}, O., {Nesvorn{\'y}}, D., \& {Lambrechts}, M. 2018,
  \aap, 620, A157, \dodoi{10.1051/0004-6361/201833855}

\bibitem[{{Carter} {et~al.}(2012){Carter}, {Agol}, {Chaplin}, {Basu},
  {Bedding}, {Buchhave}, \& {Christensen-Dalsgaard}}]{Carter2012}
{Carter}, J.~A., {Agol}, E., {Chaplin}, W.~J., {et~al.} 2012, Science, 337,
  556, \dodoi{10.1126/science.1223269}

\bibitem[{{Cimerman} {et~al.}(2018){Cimerman}, {Kley}, \&
  {Kuiper}}]{Cimerman2018}
{Cimerman}, N.~P., {Kley}, W., \& {Kuiper}, R. 2018, \aap, 618, A169,
  \dodoi{10.1051/0004-6361/201833591}

\bibitem[{{Cimerman} \& {Rafikov}(2021)}]{Cimerman2021}
{Cimerman}, N.~P., \& {Rafikov}, R.~R. 2021, arXiv e-prints, arXiv:2108.01423.
\newblock \doarXiv{2108.01423}

\bibitem[{{Cui} {et~al.}(2021){Cui}, {Papaloizou}, \& {Szuszkiewicz}}]{Cui2021}
{Cui}, Z., {Papaloizou}, J. C.~B., \& {Szuszkiewicz}, E. 2021, arXiv e-prints,
  arXiv:2108.00792.
\newblock \doarXiv{2108.00792}

\bibitem[{{Cumming} {et~al.}(2008){Cumming}, {Butler}, {Marcy}, {Vogt},
  {Wright}, \& {Fischer}}]{Cumming2008}
{Cumming}, A., {Butler}, R.~P., {Marcy}, G.~W., {et~al.} 2008, \pasp, 120, 531,
  \dodoi{10.1086/588487}

\bibitem[{{de Val-Borro} {et~al.}(2007){de Val-Borro}, {Artymowicz},
  {D'Angelo}, \& {Peplinski}}]{deValBorro2007}
{de Val-Borro}, M., {Artymowicz}, P., {D'Angelo}, G., \& {Peplinski}, A. 2007,
  \aap, 471, 1043, \dodoi{10.1051/0004-6361:20077169}

\bibitem[{{de Val-Borro} {et~al.}(2006){de Val-Borro}, {Edgar}, {Artymowicz},
  {Ciecielag}, {Cresswell}, {D'Angelo}, {Delgado-Donate}, {Dirksen}, {Fromang},
  {Gawryszczak}, {Klahr}, {Kley}, {Lyra}, {Masset}, {Mellema}, {Nelson},
  {Paardekooper}, {Peplinski}, {Pierens}, {Plewa}, {Rice}, {Sch{\"a}fer}, \&
  {Speith}}]{deValBoro2006}
{de Val-Borro}, M., {Edgar}, R.~G., {Artymowicz}, P., {et~al.} 2006, \mnras,
  370, 529, \dodoi{10.1111/j.1365-2966.2006.10488.x}

\bibitem[{{Dempsey} {et~al.}(2020){Dempsey}, {Lee}, \&
  {Lithwick}}]{Dempsey2020}
{Dempsey}, A.~M., {Lee}, W.-K., \& {Lithwick}, Y. 2020, \apj, 891, 108,
  \dodoi{10.3847/1538-4357/ab723c}

\bibitem[{{Dipierro} {et~al.}(2015){Dipierro}, {Price}, {Laibe}, {Hirsh},
  {Cerioli}, \& {Lodato}}]{Dipierro2015}
{Dipierro}, G., {Price}, D., {Laibe}, G., {et~al.} 2015, \mnras, 453, L73,
  \dodoi{10.1093/mnrasl/slv105}

\bibitem[{{Dong} {et~al.}(2011{\natexlab{a}}){Dong}, {Rafikov}, \&
  {Stone}}]{Dong2011b}
{Dong}, R., {Rafikov}, R.~R., \& {Stone}, J.~M. 2011{\natexlab{a}}, \apj, 741,
  57, \dodoi{10.1088/0004-637X/741/1/57}

\bibitem[{{Dong} {et~al.}(2011{\natexlab{b}}){Dong}, {Rafikov}, {Stone}, \&
  {Petrovich}}]{Dong2011a}
{Dong}, R., {Rafikov}, R.~R., {Stone}, J.~M., \& {Petrovich}, C.
  2011{\natexlab{b}}, \apj, 741, 56, \dodoi{10.1088/0004-637X/741/1/56}

\bibitem[{{Dong} {et~al.}(2015){Dong}, {Zhu}, \& {Whitney}}]{Dong2015}
{Dong}, R., {Zhu}, Z., \& {Whitney}, B. 2015, \apj, 809, 93,
  \dodoi{10.1088/0004-637X/809/1/93}

\bibitem[{{Duffell}(2015)}]{Duffell2015}
{Duffell}, P.~C. 2015, \apjl, 807, L11, \dodoi{10.1088/2041-8205/807/1/L11}

\bibitem[{{Duffell}(2020)}]{Duffell2020}
---. 2020, \apj, 889, 16, \dodoi{10.3847/1538-4357/ab5b0f}

\bibitem[{{Duffell} \& {Dong}(2015)}]{DuffellDong2015}
{Duffell}, P.~C., \& {Dong}, R. 2015, \apj, 802, 42,
  \dodoi{10.1088/0004-637X/802/1/42}

\bibitem[{{Fedele} {et~al.}(2021){Fedele}, {Toci}, {Maud}, \&
  {Lodato}}]{Fedele2021}
{Fedele}, D., {Toci}, C., {Maud}, L., \& {Lodato}, G. 2021, \aap, 651, A90,
  \dodoi{10.1051/0004-6361/202141278}

\bibitem[{{Flaherty} {et~al.}(2015){Flaherty}, {Hughes}, {Rosenfeld},
  {Andrews}, {Chiang}, {Simon}, {Kerzner}, \& {Wilner}}]{Flaherty2015}
{Flaherty}, K.~M., {Hughes}, A.~M., {Rosenfeld}, K.~A., {et~al.} 2015, \apj,
  813, 99, \dodoi{10.1088/0004-637X/813/2/99}

\bibitem[{{Flaherty} {et~al.}(2018){Flaherty}, {Hughes}, {Teague}, {Simon},
  {Andrews}, \& {Wilner}}]{Flaherty2018}
{Flaherty}, K.~M., {Hughes}, A.~M., {Teague}, R., {et~al.} 2018, \apj, 856,
  117, \dodoi{10.3847/1538-4357/aab615}

\bibitem[{{Flock} {et~al.}(2015){Flock}, {Ruge}, {Dzyurkevich}, {Henning},
  {Klahr}, \& {Wolf}}]{Flock2015}
{Flock}, M., {Ruge}, J.~P., {Dzyurkevich}, N., {et~al.} 2015, \aap, 574, A68,
  \dodoi{10.1051/0004-6361/201424693}

\bibitem[{{Francis} \& {van der Marel}(2020)}]{Francis2020}
{Francis}, L., \& {van der Marel}, N. 2020, \apj, 892, 111,
  \dodoi{10.3847/1538-4357/ab7b63}

\bibitem[{{Fung} \& {Ono}(2021)}]{Fung2021}
{Fung}, J., \& {Ono}, T. 2021, \apj, 922, 13, \dodoi{10.3847/1538-4357/ac1d4e}

\bibitem[{{Goldreich} \& {Tremaine}(1980)}]{Goldreich1980}
{Goldreich}, P., \& {Tremaine}, S. 1980, \apj, 241, 425, \dodoi{10.1086/158356}

\bibitem[{{Goodman} \& {Rafikov}(2001)}]{Goodman2001}
{Goodman}, J., \& {Rafikov}, R.~R. 2001, \apj, 552, 793, \dodoi{10.1086/320572}

\bibitem[{{Haffert} {et~al.}(2019){Haffert}, {Bohn}, {de Boer}, {Snellen},
  {Brinchmann}, {Girard}, {Keller}, \& {Bacon}}]{Haffert2019}
{Haffert}, S.~Y., {Bohn}, A.~J., {de Boer}, J., {et~al.} 2019, Nature
  Astronomy, 3, 749, \dodoi{10.1038/s41550-019-0780-5}

\bibitem[{{Hallam} \& {Paardekooper}(2017)}]{Hallam2017}
{Hallam}, P.~D., \& {Paardekooper}, S.~J. 2017, \mnras, 469, 3813,
  \dodoi{10.1093/mnras/stx1073}

\bibitem[{{Hammer} {et~al.}(2021){Hammer}, {Lin}, {Kratter}, \&
  {Pinilla}}]{Hammer2021}
{Hammer}, M., {Lin}, M.-K., {Kratter}, K.~M., \& {Pinilla}, P. 2021, \mnras,
  \dodoi{10.1093/mnras/stab1079}

\bibitem[{{Hayashi}(1981)}]{Hayashi1981}
{Hayashi}, C. 1981, Progress of Theoretical Physics Supplement, 70, 35,
  \dodoi{10.1143/PTPS.70.35}

\bibitem[{{Hunter}(2007)}]{Hunter2007}
{Hunter}, J.~D. 2007, Computing in Science and Engineering, 9, 90,
  \dodoi{10.1109/MCSE.2007.55}

\bibitem[{{Isella} {et~al.}(2018){Isella}, {Huang}, {Andrews}, {Dullemond},
  {Birnstiel}, {Zhang}, {Zhu}, {Guzm{\'a}n}, {P{\'e}rez}, {Bai}, {Benisty},
  {Carpenter}, {Ricci}, \& {Wilner}}]{Isella2018}
{Isella}, A., {Huang}, J., {Andrews}, S.~M., {et~al.} 2018, \apjl, 869, L49,
  \dodoi{10.3847/2041-8213/aaf747}

\bibitem[{{Kanagawa} {et~al.}(2016){Kanagawa}, {Muto}, {Tanaka}, {Tanigawa},
  {Takeuchi}, {Tsukagoshi}, \& {Momose}}]{Kanagawa2016}
{Kanagawa}, K.~D., {Muto}, T., {Tanaka}, H., {et~al.} 2016, \pasj, 68, 43,
  \dodoi{10.1093/pasj/psw037}

\bibitem[{{Kanagawa} \& {Szuszkiewicz}(2020)}]{Kanagawa2020}
{Kanagawa}, K.~D., \& {Szuszkiewicz}, E. 2020, \apj, 894, 59,
  \dodoi{10.3847/1538-4357/ab862f}

\bibitem[{{Kanagawa} {et~al.}(2017){Kanagawa}, {Tanaka}, {Muto}, \&
  {Tanigawa}}]{Kanagawa2017}
{Kanagawa}, K.~D., {Tanaka}, H., {Muto}, T., \& {Tanigawa}, T. 2017, \pasj, 69,
  97, \dodoi{10.1093/pasj/psx114}

\bibitem[{{Kanagawa} {et~al.}(2015){Kanagawa}, {Tanaka}, {Muto}, {Tanigawa}, \&
  {Takeuchi}}]{Kanagawa2015}
{Kanagawa}, K.~D., {Tanaka}, H., {Muto}, T., {Tanigawa}, T., \& {Takeuchi}, T.
  2015, \mnras, 448, 994, \dodoi{10.1093/mnras/stv025}

\bibitem[{{Kanagawa} {et~al.}(2018){Kanagawa}, {Tanaka}, \&
  {Szuszkiewicz}}]{Kanagawa2018}
{Kanagawa}, K.~D., {Tanaka}, H., \& {Szuszkiewicz}, E. 2018, \apj, 861, 140,
  \dodoi{10.3847/1538-4357/aac8d9}

\bibitem[{{Keppler} {et~al.}(2018){Keppler}, {Benisty}, {M{\"u}ller},
  {Henning}, {van Boekel}, {Cantalloube}, {Ginski}, {van Holstein}, {Maire},
  {Pohl}, {Samland}, {Avenhaus}, {Baudino}, {Boccaletti}, {de Boer},
  {Bonnefoy}, {Chauvin}, {Desidera}, {Langlois}, {Lazzoni}, {Marleau},
  {Mordasini}, {Pawellek}, {Stolker}, {Vigan}, {Zurlo}, {Birnstiel},
  {Brandner}, {Feldt}, {Flock}, {Girard}, {Gratton}, {Hagelberg}, {Isella},
  {Janson}, {Juhasz}, {Kemmer}, {Kral}, {Lagrange}, {Launhardt}, {Matter},
  {M{\'e}nard}, {Milli}, {Molli{\`e}re}, {Olofsson}, {P{\'e}rez}, {Pinilla},
  {Pinte}, {Quanz}, {Schmidt}, {Udry}, {Wahhaj}, {Williams}, {Buenzli},
  {Cudel}, {Dominik}, {Galicher}, {Kasper}, {Lannier}, {Mesa}, {Mouillet},
  {Peretti}, {Perrot}, {Salter}, {Sissa}, {Wildi}, {Abe}, {Antichi},
  {Augereau}, {Baruffolo}, {Baudoz}, {Bazzon}, {Beuzit}, {Blanchard}, {Brems},
  {Buey}, {De Caprio}, {Carbillet}, {Carle}, {Cascone}, {Cheetham}, {Claudi},
  {Costille}, {Delboulb{\'e}}, {Dohlen}, {Fantinel}, {Feautrier}, {Fusco},
  {Giro}, {Gluck}, {Gry}, {Hubin}, {Hugot}, {Jaquet}, {Le Mignant}, {Llored},
  {Madec}, {Magnard}, {Martinez}, {Maurel}, {Meyer}, {M{\"o}ller-Nilsson},
  {Moulin}, {Mugnier}, {Orign{\'e}}, {Pavlov}, {Perret}, {Petit}, {Pragt},
  {Puget}, {Rabou}, {Ramos}, {Rigal}, {Rochat}, {Roelfsema}, {Rousset}, {Roux},
  {Salasnich}, {Sauvage}, {Sevin}, {Soenke}, {Stadler}, {Suarez}, {Turatto}, \&
  {Weber}}]{Keppler2018}
{Keppler}, M., {Benisty}, M., {M{\"u}ller}, A., {et~al.} 2018, \aap, 617, A44,
  \dodoi{10.1051/0004-6361/201832957}

\bibitem[{{Kokubo} \& {Ida}(1998)}]{Kokubo98}
{Kokubo}, E., \& {Ida}, S. 1998, \icarus, 131, 171,
  \dodoi{10.1006/icar.1997.5840}

\bibitem[{{Krapp} {et~al.}(2018){Krapp}, {Gressel}, {Ben{\'\i}tez-Llambay},
  {Downes}, {Mohandas}, \& {Pessah}}]{Krapp2018}
{Krapp}, L., {Gressel}, O., {Ben{\'\i}tez-Llambay}, P., {et~al.} 2018, \apj,
  865, 105, \dodoi{10.3847/1538-4357/aadcf0}

\bibitem[{{Les} \& {Lin}(2015)}]{Les2015}
{Les}, R., \& {Lin}, M.-K. 2015, \mnras, 450, 1503,
  \dodoi{10.1093/mnras/stv712}

\bibitem[{{Li} {et~al.}(2001){Li}, {Colgate}, {Wendroff}, \& {Liska}}]{Li2001}
{Li}, H., {Colgate}, S.~A., {Wendroff}, B., \& {Liska}, R. 2001, \apj, 551,
  874, \dodoi{10.1086/320241}

\bibitem[{{Li} {et~al.}(2000){Li}, {Finn}, {Lovelace}, \& {Colgate}}]{Li2000}
{Li}, H., {Finn}, J.~M., {Lovelace}, R.~V.~E., \& {Colgate}, S.~A. 2000, \apj,
  533, 1023, \dodoi{10.1086/308693}

\bibitem[{{Lin} \& {Papaloizou}(1986)}]{LinDNC1986}
{Lin}, D.~N.~C., \& {Papaloizou}, J. 1986, \apj, 309, 846,
  \dodoi{10.1086/164653}

\bibitem[{{Lin}(2012)}]{Lin2012}
{Lin}, M.-K. 2012, \apj, 754, 21, \dodoi{10.1088/0004-637X/754/1/21}

\bibitem[{{Lissauer} {et~al.}(2011{\natexlab{a}}){Lissauer}, {Ragozzine},
  {Fabrycky}, {Steffen}, {Ford}, {Jenkins}, {Shporer}, {Holman}, {Rowe},
  {Quintana}, {Batalha}, {Borucki}, {Bryson}, {Caldwell}, {Carter}, {Ciardi},
  {Dunham}, {Fortney}, {Gautier}, {Howell}, {Koch}, {Latham}, {Marcy},
  {Morehead}, \& {Sasselov}}]{Lissauer2011a}
{Lissauer}, J.~J., {Ragozzine}, D., {Fabrycky}, D.~C., {et~al.}
  2011{\natexlab{a}}, \apjs, 197, 8, \dodoi{10.1088/0067-0049/197/1/8}

\bibitem[{{Lissauer} {et~al.}(2011{\natexlab{b}}){Lissauer}, {Fabrycky},
  {Ford}, {Borucki}, {Fressin}, {Marcy}, {Orosz}, {Rowe}, {Torres}, {Welsh},
  {Batalha}, {Bryson}, {Buchhave}, {Caldwell}, {Carter}, {Charbonneau},
  {Christiansen}, {Cochran}, {Desert}, {Dunham}, {Fanelli}, {Fortney},
  {Gautier}, {Geary}, {Gilliland}, {Haas}, {Hall}, {Holman}, {Koch}, {Latham},
  {Lopez}, {McCauliff}, {Miller}, {Morehead}, {Quintana}, {Ragozzine},
  {Sasselov}, {Short}, \& {Steffen}}]{Lissauer2011b}
{Lissauer}, J.~J., {Fabrycky}, D.~C., {Ford}, E.~B., {et~al.}
  2011{\natexlab{b}}, \nat, 470, 53, \dodoi{10.1038/nature09760}

\bibitem[{{Lissauer} {et~al.}(2014){Lissauer}, {Marcy}, {Bryson}, {Rowe},
  {Jontof-Hutter}, {Agol}, {Borucki}, {Carter}, {Ford}, {Gilliland}, {Kolbl},
  {Star}, {Steffen}, \& {Torres}}]{Lissauer2014}
{Lissauer}, J.~J., {Marcy}, G.~W., {Bryson}, S.~T., {et~al.} 2014, \apj, 784,
  44, \dodoi{10.1088/0004-637X/784/1/44}

\bibitem[{{Lodato} {et~al.}(2019){Lodato}, {Dipierro}, {Ragusa}, {Long},
  {Herczeg}, {Pascucci}, {Pinilla}, {Manara}, {Tazzari}, {Liu}, {Mulders},
  {Harsono}, {Boehler}, {M{\'e}nard}, {Johnstone}, {Salyk}, {van der Plas},
  {Cabrit}, {Edwards}, {Fischer}, {Hendler}, {Nisini}, {Rigliaco}, {Avenhaus},
  {Banzatti}, \& {Gully-Santiago}}]{Lodato2019}
{Lodato}, G., {Dipierro}, G., {Ragusa}, E., {et~al.} 2019, \mnras, 486, 453,
  \dodoi{10.1093/mnras/stz913}

\bibitem[{{Long} {et~al.}(2018){Long}, {Pinilla}, {Herczeg}, {Harsono},
  {Dipierro}, {Pascucci}, {Hendler}, {Tazzari}, {Ragusa}, {Salyk}, {Edwards},
  {Lodato}, {van de Plas}, {Johnstone}, {Liu}, {Boehler}, {Cabrit}, {Manara},
  {Menard}, {Mulders}, {Nisini}, {Fischer}, {Rigliaco}, {Banzatti}, {Avenhaus},
  \& {Gully-Santiago}}]{Long2018}
{Long}, F., {Pinilla}, P., {Herczeg}, G.~J., {et~al.} 2018, \apj, 869, 17,
  \dodoi{10.3847/1538-4357/aae8e1}

\bibitem[{{Lovelace} {et~al.}(1999){Lovelace}, {Li}, {Colgate}, \&
  {Nelson}}]{Lovelace1999}
{Lovelace}, R.~V.~E., {Li}, H., {Colgate}, S.~A., \& {Nelson}, A.~F. 1999,
  \apj, 513, 805, \dodoi{10.1086/306900}

\bibitem[{{Malhotra}(1993)}]{Malhotra1993}
{Malhotra}, R. 1993, \nat, 365, 819, \dodoi{10.1038/365819a0}

\bibitem[{{Masset}(2000)}]{Masset2000}
{Masset}, F. 2000, \aaps, 141, 165, \dodoi{10.1051/aas:2000116}

\bibitem[{{Mayor} {et~al.}(2011){Mayor}, {Marmier}, {Lovis}, {Udry},
  {S{\'e}gransan}, {Pepe}, {Benz}, {Bertaux}, {Bouchy}, {Dumusque}, {Lo Curto},
  {Mordasini}, {Queloz}, \& {Santos}}]{Mayor2011}
{Mayor}, M., {Marmier}, M., {Lovis}, C., {et~al.} 2011, arXiv e-prints,
  arXiv:1109.2497.
\newblock \doarXiv{1109.2497}

\bibitem[{{Miranda} \& {Rafikov}(2020)}]{Miranda2020}
{Miranda}, R., \& {Rafikov}, R.~R. 2020, \apj, 904, 121,
  \dodoi{10.3847/1538-4357/abbee7}

\bibitem[{{Montesinos} {et~al.}(2020){Montesinos}, {Garrido-Deutelmoser},
  {Olofsson}, {Giuppone}, {Cuadra}, {Bayo}, {Sucerquia}, \&
  {Cuello}}]{Montesinos2020}
{Montesinos}, M., {Garrido-Deutelmoser}, J., {Olofsson}, J., {et~al.} 2020,
  \aap, 642, A224, \dodoi{10.1051/0004-6361/202038758}

\bibitem[{{M{\"u}ller} {et~al.}(2018){M{\"u}ller}, {Keppler}, {Henning},
  {Samland}, {Chauvin}, {Beust}, {Maire}, {Molaverdikhani}, {van Boekel},
  {Benisty}, {Boccaletti}, {Bonnefoy}, {Cantalloube}, {Charnay}, {Baudino},
  {Gennaro}, {Long}, {Cheetham}, {Desidera}, {Feldt}, {Fusco}, {Girard},
  {Gratton}, {Hagelberg}, {Janson}, {Lagrange}, {Langlois}, {Lazzoni}, {Ligi},
  {M{\'e}nard}, {Mesa}, {Meyer}, {Molli{\`e}re}, {Mordasini}, {Moulin},
  {Pavlov}, {Pawellek}, {Quanz}, {Ramos}, {Rouan}, {Sissa}, {Stadler}, {Vigan},
  {Wahhaj}, {Weber}, \& {Zurlo}}]{Muller2018}
{M{\"u}ller}, A., {Keppler}, M., {Henning}, T., {et~al.} 2018, \aap, 617, L2,
  \dodoi{10.1051/0004-6361/201833584}

\bibitem[{{{\"O}berg} {et~al.}(2021){{\"O}berg}, {Guzm{\'a}n}, {Walsh},
  {Aikawa}, {Bergin}, {Law}, {Loomis}, {Alarc{\'o}n}, {Andrews}, {Bae},
  {Bergner}, {Boehler}, {Booth}, {Bosman}, {Calahan}, {Cataldi}, {Cleeves},
  {Czekala}, {Furuya}, {Huang}, {Ilee}, {Kurtovic}, {Le Gal}, {Liu}, {Long},
  {M{\'e}nard}, {Nomura}, {P{\'e}rez}, {Qi}, {Schwarz}, {Sierra}, {Teague},
  {Tsukagoshi}, {Yamato}, {van't Hoff}, {Waggoner}, {Wilner}, \&
  {Zhang}}]{MAPS_I}
{{\"O}berg}, K.~I., {Guzm{\'a}n}, V.~V., {Walsh}, C., {et~al.} 2021, \apjs,
  257, 1, \dodoi{10.3847/1538-4365/ac1432}

\bibitem[{{Okuzumi} {et~al.}(2016){Okuzumi}, {Momose}, {Sirono}, {Kobayashi},
  \& {Tanaka}}]{Okuzumi2016}
{Okuzumi}, S., {Momose}, M., {Sirono}, S.-i., {Kobayashi}, H., \& {Tanaka}, H.
  2016, \apj, 821, 82, \dodoi{10.3847/0004-637X/821/2/82}

\bibitem[{{Ono} {et~al.}(2016){Ono}, {Muto}, {Takeuchi}, \& {Nomura}}]{Ono2016}
{Ono}, T., {Muto}, T., {Takeuchi}, T., \& {Nomura}, H. 2016, \apj, 823, 84,
  \dodoi{10.3847/0004-637X/823/2/84}

\bibitem[{{P{\'e}rez} {et~al.}(2020){P{\'e}rez}, {Casassus}, {Hales}, {Marino},
  {Cheetham}, {Zurlo}, {Cieza}, {Dong}, {Alarc{\'o}n}, {Ben{\'\i}tez-Llambay},
  {Fomalont}, \& {Avenhaus}}]{Perez2020}
{P{\'e}rez}, S., {Casassus}, S., {Hales}, A., {et~al.} 2020, \apjl, 889, L24,
  \dodoi{10.3847/2041-8213/ab6b2b}

\bibitem[{{Petigura} {et~al.}(2013){Petigura}, {Howard}, \&
  {Marcy}}]{Petigura2013}
{Petigura}, E.~A., {Howard}, A.~W., \& {Marcy}, G.~W. 2013, Proceedings of the
  National Academy of Science, 110, 19273, \dodoi{10.1073/pnas.1319909110}

\bibitem[{{Petrovich} \& {Rafikov}(2012)}]{PR2012}
{Petrovich}, C., \& {Rafikov}, R.~R. 2012, \apj, 758, 33,
  \dodoi{10.1088/0004-637X/758/1/33}

\bibitem[{{Pierens}(2021)}]{Pierens2021}
{Pierens}, A. 2021, \mnras, 504, 4522, \dodoi{10.1093/mnras/stab183}

\bibitem[{{Pinilla} {et~al.}(2012){Pinilla}, {Benisty}, \&
  {Birnstiel}}]{Pinilla2012}
{Pinilla}, P., {Benisty}, M., \& {Birnstiel}, T. 2012, \aap, 545, A81,
  \dodoi{10.1051/0004-6361/201219315}

\bibitem[{{Pu} \& {Wu}(2015)}]{PuWu2015}
{Pu}, B., \& {Wu}, Y. 2015, \apj, 807, 44, \dodoi{10.1088/0004-637X/807/1/44}

\bibitem[{{Rafikov}(2002)}]{Rafivok2002a}
{Rafikov}, R.~R. 2002, \apj, 569, 997, \dodoi{10.1086/339399}

\bibitem[{{Rafikov} \& {Petrovich}(2012)}]{RP2012}
{Rafikov}, R.~R., \& {Petrovich}, C. 2012, \apj, 747, 24,
  \dodoi{10.1088/0004-637X/747/1/24}

\bibitem[{{Rodenkirch} {et~al.}(2021){Rodenkirch}, {Rometsch}, {Dullemond},
  {Weber}, \& {Kley}}]{Rodenkirch2021}
{Rodenkirch}, P.~J., {Rometsch}, T., {Dullemond}, C.~P., {Weber}, P., \&
  {Kley}, W. 2021, \aap, 647, A174, \dodoi{10.1051/0004-6361/202038484}

\bibitem[{{Rometsch} {et~al.}(2021){Rometsch}, {Ziampras}, {Kley}, \&
  {B{\'e}thune}}]{Rometsch2021}
{Rometsch}, T., {Ziampras}, A., {Kley}, W., \& {B{\'e}thune}, W. 2021, \aap,
  656, A130, \dodoi{10.1051/0004-6361/202142105}

\bibitem[{{Rowe} {et~al.}(2014){Rowe}, {Bryson}, {Marcy}, {Lissauer},
  {Jontof-Hutter}, {Mullally}, {Gilliland}, {Issacson}, {Ford}, {Howell},
  {Borucki}, {Haas}, {Huber}, {Steffen}, {Thompson}, {Quintana}, {Barclay},
  {Still}, {Fortney}, {Gautier}, {Hunter}, {Caldwell}, {Ciardi}, {Devore},
  {Cochran}, {Jenkins}, {Agol}, {Carter}, \& {Geary}}]{Rowe2014}
{Rowe}, J.~F., {Bryson}, S.~T., {Marcy}, G.~W., {et~al.} 2014, \apj, 784, 45,
  \dodoi{10.1088/0004-637X/784/1/45}

\bibitem[{{Shakura} \& {Sunyaev}(1973)}]{Shakura1973}
{Shakura}, N.~I., \& {Sunyaev}, R.~A. 1973, \aap, 500, 33

\bibitem[{{Surville} \& {Barge}(2015)}]{Surville2015}
{Surville}, C., \& {Barge}, P. 2015, \aap, 579, A100,
  \dodoi{10.1051/0004-6361/201424663}

\bibitem[{{Takahashi} \& {Inutsuka}(2016)}]{Takahashi2016}
{Takahashi}, S.~Z., \& {Inutsuka}, S.-i. 2016, \aj, 152, 184,
  \dodoi{10.3847/0004-6256/152/6/184}

\bibitem[{{Tsiganis} {et~al.}(2005){Tsiganis}, {Gomes}, {Morbidelli}, \&
  {Levison}}]{Tsiganis2005}
{Tsiganis}, K., {Gomes}, R., {Morbidelli}, A., \& {Levison}, H.~F. 2005, \nat,
  435, 459, \dodoi{10.1038/nature03539}

\bibitem[{{Ueda} {et~al.}(2021){Ueda}, {Flock}, \& {Birnstiel}}]{Takahiro2021}
{Ueda}, T., {Flock}, M., \& {Birnstiel}, T. 2021, \apjl, 914, L38,
  \dodoi{10.3847/2041-8213/ac0631}

\bibitem[{{van der Marel} \& {Mulders}(2021)}]{MarelMulders}
{van der Marel}, N., \& {Mulders}, G.~D. 2021, \aj, 162, 28,
  \dodoi{10.3847/1538-3881/ac0255}

\bibitem[{{van der Walt} {et~al.}(2011){van der Walt}, {Colbert}, \&
  {Varoquaux}}]{vanderWalt2011}
{van der Walt}, S., {Colbert}, S.~C., \& {Varoquaux}, G. 2011, Computing in
  Science and Engineering, 13, 22, \dodoi{10.1109/MCSE.2011.37}

\bibitem[{{Wagner} {et~al.}(2018){Wagner}, {Follete}, {Close}, {Apai}, {Gibbs},
  {Keppler}, {M{\"u}ller}, {Henning}, {Kasper}, {Wu}, {Long}, {Males},
  {Morzinski}, \& {McClure}}]{Wagner2018}
{Wagner}, K., {Follete}, K.~B., {Close}, L.~M., {et~al.} 2018, \apjl, 863, L8,
  \dodoi{10.3847/2041-8213/aad695}

\bibitem[{{Yee} {et~al.}(2021){Yee}, {Tamayo}, {Hadden}, \& {Winn}}]{Yee2021}
{Yee}, S.~W., {Tamayo}, D., {Hadden}, S., \& {Winn}, J.~N. 2021, \aj, 162, 55,
  \dodoi{10.3847/1538-3881/ac00a9}

\bibitem[{{Zhang} {et~al.}(2015){Zhang}, {Blake}, \& {Bergin}}]{Zhang2015}
{Zhang}, K., {Blake}, G.~A., \& {Bergin}, E.~A. 2015, \apjl, 806, L7,
  \dodoi{10.1088/2041-8205/806/1/L7}

\bibitem[{{Zhang} {et~al.}(2021){Zhang}, {Booth}, {Law}, {Bosman}, {Schwarz},
  {Bergin}, {{\"O}berg}, {Andrews}, {Guzm{\'a}n}, {Walsh}, {Qi}, {van't Hoff},
  {Long}, {Wilner}, {Huang}, {Czekala}, {Ilee}, {Cataldi}, {Bergner}, {Aikawa},
  {Teague}, {Bae}, {Loomis}, {Calahan}, {Alarc{\'o}n}, {M{\'e}nard}, {Le Gal},
  {Sierra}, {Yamato}, {Nomura}, {Tsukagoshi}, {P{\'e}rez}, {Trapman}, {Liu}, \&
  {Furuya}}]{zhang2021}
{Zhang}, K., {Booth}, A.~S., {Law}, C.~J., {et~al.} 2021, \apjs, 257, 5,
  \dodoi{10.3847/1538-4365/ac1580}

\bibitem[{{Zhang} \& {Zhu}(2020)}]{Zhang&Zhu2020}
{Zhang}, S., \& {Zhu}, Z. 2020, \mnras, 493, 2287,
  \dodoi{10.1093/mnras/staa404}

\bibitem[{{Zhang} {et~al.}(2018){Zhang}, {Zhu}, {Huang}, {Guzm{\'a}n},
  {Andrews}, {Birnstiel}, {Dullemond}, {Carpenter}, {Isella}, {P{\'e}rez},
  {Benisty}, {Wilner}, {Baruteau}, {Bai}, \& {Ricci}}]{Zhang2018}
{Zhang}, S., {Zhu}, Z., {Huang}, J., {et~al.} 2018, \apjl, 869, L47,
  \dodoi{10.3847/2041-8213/aaf744}

\bibitem[{{Zhu} {et~al.}(2018){Zhu}, {Petrovich}, {Wu}, {Dong}, \&
  {Xie}}]{Zhu2018}
{Zhu}, W., {Petrovich}, C., {Wu}, Y., {Dong}, S., \& {Xie}, J. 2018, \apj, 860,
  101, \dodoi{10.3847/1538-4357/aac6d5}

\bibitem[{{Zhu} {et~al.}(2011){Zhu}, {Nelson}, {Hartmann}, {Espaillat}, \&
  {Calvet}}]{Zhu2011}
{Zhu}, Z., {Nelson}, R.~P., {Hartmann}, L., {Espaillat}, C., \& {Calvet}, N.
  2011, \apj, 729, 47, \dodoi{10.1088/0004-637X/729/1/47}

\end{thebibliography}
\bibliographystyle{aasjournal}


\end{document}